\documentclass[journal,12pt,onecolumn,draftclsnofoot,]{IEEEtran}

\usepackage[T1]{fontenc}
\usepackage{color}
\usepackage{varioref}
\usepackage{textcomp}
\usepackage{amsthm}
\usepackage{amsmath}
\usepackage{amssymb}
\usepackage{graphicx}
\usepackage{setspace}
\usepackage{esint}
\usepackage{algorithm}
\usepackage{algpseudocode}
\usepackage{pifont}
\usepackage{amsmath}

\usepackage{enumitem}
\usepackage{cite}
\usepackage[utf8]{inputenc}
\usepackage{array}
\usepackage{wrapfig}
\usepackage{multirow}
\usepackage{tabu}
\makeatletter

\newcommand{\Rmnum}[1]{\expandafter\@slowromancap\romannumeral #1@}
\makeatother

\newtheorem{theorem}{Theorem}

\newtheorem{remark}{Remark}
\newtheorem{definition}{Definition}
\newtheorem{corollary}{Corollary}

\newtheorem{thm}{\protect\theoremname}
\newtheorem{prop}[thm]{\protect\propositionname}
\providecommand{\propositionname}{Proposition}

\ifCLASSINFOpdf

\else

\fi

\usepackage[T1]{fontenc}
\usepackage{etoolbox}

\makeatletter
\patchcmd{\maketitle}{\@fnsymbol}{\@alph}{}{}  
\makeatother

\title{Cache-Aided Content Delivery over Erasure Broadcast Channels}
\author{
  Mohammad Mohammadi Amiri and
  Deniz~G\"und\"uz
}
\date{}

\begin{document}

\maketitle

\begin{abstract}
A cache-aided broadcast network is studied, in which a server delivers contents to a group of receivers over a packet erasure broadcast channel (BC). The receivers are divided into two sets with regards to their channel qualities: the \textit{weak} and \textit{strong} receivers, where all the weak receivers have statistically worse channel qualities than all the strong receivers. The weak receivers, in order to compensate for the high erasure probability they encounter over the channel, are equipped with cache memories of equal size, while the receivers in the strong set have no caches. Data can be pre-delivered to weak receivers' caches over the off-peak traffic period before the receivers reveal their demands. Allowing arbitrary erasure probabilities for the weak and strong receivers, a joint caching and channel coding scheme, which divides each file into several subfiles, and applies a different caching and delivery scheme for each subfile, is proposed. It is shown that all the receivers, even those without any cache memories, benefit from the presence of caches across the network. An information theoretic trade-off between the cache size and the achievable rate is formulated. It is shown that the proposed scheme improves upon the state-of-the-art in terms of the achievable trade-off.   
\end{abstract}


\section{Introduction}\label{Intro}
Wireless content caching is a promising technique to flatten the traffic over the backhaul network by shifting part of the traffic from peak to off-peak periods \cite{DowdyCaching,AlmerothCacing,MeyersonMunagalaPlotkin,BaevRajaramanSwamy,BorstGuptaWalid}. Video-on-demand services for mobile users would particularly benefit from content caching as a few highly popular files are requested by a large number of users over a relatively short time period. Popular contents that are likely to be requested by a majority of the users can be proactively cached at the network edge during a period of low network traffic, known as the \textit{placement phase}. The \textit{delivery phase} is performed during a peak traffic period, when the users reveal their demands, and the cached contents can be exploited to reduce both the load over the backhaul links and the latency in delivery \cite{MaddahAliCentralized}.

A coded proactive content caching and delivery scheme has been proposed by Maddah-Ali and Niesen in \cite{MaddahAliCentralized,MaddahAliDecentralized}, where they consider a library of same-size files to be delivered over a noiseless broadcast channel (BC), while the receivers are equipped with cache memories of equal size. They identify a trade-off between the cache size and the minimum rate required during the delivery phase to serve all the receivers for all demand combinations, and show that coding can significantly reduce the required delivery rate. Several improved coded caching schemes and information theoretic performance bounds have been introduced since then \cite{ZhiChenXOR,MohammadDenizTCom,MohammadQianDenizITW,YuMaddahAliAvestimehrExact,TianComputerAidedLongerVersion,SenguptaCaching,GhasemiCachingLowerBound,GastparNewConverseBound}. Coded caching has since been extended to various network settings, including device-to-device caching \cite{GregoryDtoD,JiCaireMolischDtoD}, online cache placement \cite{PedarsaniOnlineCaching}, files with non-uniform popularities \cite{NiesenNonuniform,ZhangCachingArbitraryDemands} and distinct lengths \cite{ZhangDistinctFileSizes}, users with non-uniform cache sizes \cite{WangHeterogenous,MohammadQianDenizDistinctCacheSizes}, multi-layer caching \cite{KaramchandaniHierarchical}, and caching by users with different distortion requirements \cite{QianDenizLossy,ElzaDistortionMemoryTradeoff,TimoDistortionCaching}.          

In contrast to the setting introduced in \cite{MaddahAliCentralized}, a noisy channel is considered for the delivery phase in \cite{EliaCSITFeddbackJournal,GhorbelErasureCacheFeedback,TimoJointCacheChannelCoding,ShirinErasureChannelJournal,HuangWirelessFadingChannel,MaddahAliInterferenceJournal,NaderializadehMaddahAliInterference,PetrosEliaTopological}. Here, we follow the model considered in \cite{ShirinErasureChannelJournal}, and assume that the delivery phase takes place over a memoryless packet erasure BC, which models a packetized communication system, where each packet is separately channel coded against errors at the physical layer, so that a packet either arrives at a receiver correctly, or is lost. Communication over Internet is usually modeled as a packet erasure channel. The receivers in the system are grouped into two disjoint sets of weak and strong receivers. All the weak receivers are assumed to have statistically worse channels than the strong receivers, while the users in each set can have arbitrary erasure probabilities. To compensate for the worse channel quality, each weak receiver is equipped with a cache memory of equal size. We consider the case when the number of receivers is not greater than the number of files in the library. Assuming equal-rate files in the library, we derive a trade-off between the size of the caches provided to the weak receivers and the rate of the files, for which any demand combination can be reliably satisfied over the erasure BC. The proposed scheme exploits a novel file subpacketization, and performs a different caching and content delivery scheme to deliver different subpackets over the channel. Moreover, the delivery of the contents to the weak and strong receivers are coupled through the use of a joint encoding scheme to maximally benefit from the available cache memories. We show that, when specified to the homogeneous scenario considered in \cite{ShirinErasureChannelJournal}, where the receivers in the same set have the same erasure probability, the proposed scheme outperforms the one in \cite{ShirinErasureChannelJournal}.            

The rest of this paper is organized as follows. We introduce the system model in Section \ref{SystemModel}. Main results are summarized and compared with the state-of-the-art in Section \ref{AchievablePairs}. The proposed scheme is elaborated and analyzed in Section \ref{SCCHeterogeneousHomogeneous}. We conclude the paper in Section \ref{Conc}.

\textit{Notations:} For two integers $i \le j$, the set $\left\{ i, i+1, ..., j \right\}$ is denoted by $\left[ i:j \right]$, and the set $[1:i]$ is shortly denoted by $[i]$. For two sets $\mathcal{Q}$ and $\mathcal{P}$, $\mathcal{Q} \backslash \mathcal{P}$ is the set of elements in $\mathcal{Q}$ that do not belong to $\mathcal{P}$. Notation $\left| \cdot \right|$ represents the cardinality of a set, or the length of a vector. Notation $\oplus$ refers to bitwise XOR operation; and finally, $\binom{j}{i}$ returns the binomial coefficient ``$j$ choose $i$''.

\begin{figure}[!t]
\centering
\includegraphics[scale=0.6]{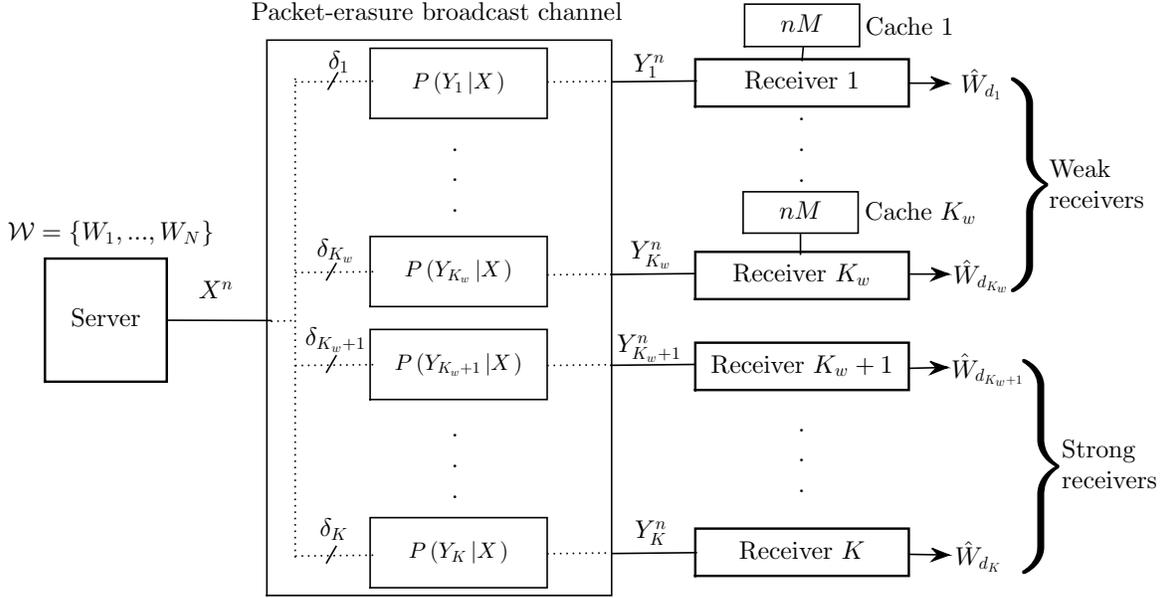}
\caption{Cache-aided packet erasure BC. The first $K_w$ receivers have statically worse channels than the last $K_s$ receivers, but each of them is equipped with a cache of normalized size $M$.} 
\label{System_Model}
\end{figure}

\section{System Model and Preliminaries}\label{SystemModel}
We consider a server with a library of $N$ popular files $\mathbf W \buildrel \Delta \over = \left( W_1, ..., W_N \right)$. Each file $W_f$ is distributed uniformly over the set $\left[ {\left\lceil {{2^{nR}}} \right\rceil } \right]$, $\forall f \in \left[ N \right]$, where $R$ denotes the rate of each file, and $n$ is the number of channel uses during the delivery phase. Receiver $k$'s demand is represented by $d_k$, where $d_k \in \left[ N \right]$, $\forall k \in \left[ K \right]$. The server delivers all the requests $W_{d_1}, ..., W_{d_K}$ to their corresponding receivers simultaneously over a BC.

Following \cite{ShirinErasureChannelJournal}, the channel between the server and the receivers is modeled as a memoryless packet erasure BC. For each channel use, the server transmits an $F$-bit codeword from the alphabet $\mathcal X \buildrel \Delta \over = \left\{ 0,1 \right\}^F$, and the output alphabet at each receiver is $\mathcal Y \buildrel \Delta \over = \mathcal X \cup \left\{ \Delta  \right\}$, where the erasure symbol $\Delta$ corresponds to a packet that is not received at the receiver. Receiver $k$, for $k \in \left[ K \right]$, receives the transmitted codeword $x \in \mathcal X$ correctly with probability $1-\delta_k$, and the erasure symbol $\Delta$ with probability $\delta_k$. Thus, given the transmitted codeword $x \in \mathcal X$, receiver $k \in [K]$ observes the output $y_k \in \mathcal Y$ with the conditional probability 
\begin{equation}\label{MarginalTransition} 
P \left( {{Y_k} = {y_k}\left| {X = x} \right.} \right) = 
\begin{cases}
1-\delta_k, &\mbox{if $y_k=x$},\\
\delta_k, &\mbox{if $y_k=\Delta$}.
\end{cases}
\end{equation}

Two disjoint sets of receivers, \textit{weak} and \textit{strong} receivers, are considered, grouped according to the erasure probabilities of their channels. These groups may model users located in areas with relatively bad and good network coverage, respectively. We assume that the channel condition of each strong receiver is statistically better than that of each weak receiver; that is, the erasure probability of a strong receiver is lower than any weak receiver. Without loss of generality, we enumerate the receivers in the order of improving channel quality, that is, we have $\delta_1 \ge \delta_2 \ge \cdots \ge \delta_K$. We denote the set of erasure probabilities by $\boldsymbol{\delta} \buildrel \Delta \over = \left\{ \delta_1, \delta_2, ..., \delta_K \right\}$. We denote the first $K_w$ receivers as the \textit{weak receivers}, and the next $K_s = K - K_w$ receivers as the \textit{strong receivers}. To compensate for their poorer channel quality, each weak receiver is equipped with a cache memory of size $nM$ bits, as depicted in Fig. \ref{System_Model}. The special case in which all the receivers in the same set have the same erasure probability; that is, all the weak receivers have erasure probability $\delta_w$, and all the strong receivers have erasure probability $\delta_s$, with $\delta_s < \delta_w$, is called the \textit{homogeneous scenario}. The set of erasure probabilities for the homogeneous scenario is represented by ${\boldsymbol{\delta}}_{ws}$.

Content delivery is performed in two phases \cite{MaddahAliCentralized}. It starts with the placement phase which takes place during the off-peak traffic period, and the caches of the weak receivers are filled without the knowledge of their future demands. Thus, only the weak receivers take part in the placement phase, and the contents of the cache of receiver $k$, for $k \in \left[ K_w \right]$, at the end of this phase is denoted by $Z_k$. Note that, $Z_k$ does not depend on any specific demand combination, and it is instead a function of the library $\mathbf{W}$. The caching function for receiver $k \in \left[ K_w \right]$ is given by
\begin{equation}\label{CachingFunction} 
{\phi _k}: {\left[ {\left\lceil {{2^{nR}}} \right\rceil } \right]^N} \to {\left[ \left\lfloor {{2^{nM}}} \right\rfloor  \right]},
\end{equation}
which maps the entire library to the cache content $Z_k$, i.e., $Z_k={\phi _k}\left( \mathbf{W} \right)$. It is to be noted that, since the placement phase is performed over a low-congestion period, it is assumed that no erasure occurs during this phase. 

The delivery phase follows once the demands of the receivers are revealed to the server, which transmits a length-$n$ codeword $X^n$ over the erasure BC. For a demand vector $\mathbf{d} \buildrel \Delta \over = (d_1, ..., d_K)$, a coded delivery function 
\begin{equation}\label{DeliveryFunction} 
\psi :{\left[ {\left\lceil {{2^{nR}}} \right\rceil } \right]^N} \times \left[N\right]^K \to \mathcal X^n
\end{equation}
generates a common message $X^n$ as a function of the entire library and the receiver demands, i.e., $X^n=\psi \left( \mathbf{W},\mathbf{d} \right)$, to be delivered over the erasure BC. Each receiver $k \in \left[K\right]$ observes $Y_k^n$ according to \eqref{MarginalTransition}. Each weak receiver $k \in \left[K_w\right]$ tries to decode $W_{d_k}$ from its channel output $Y_k^n$ along with the content available locally in its cache and the demand vector $\textbf{d}$, utilizing the decoding function
\begin{equation}\label{DecodingFunctionWeak} 
{\mu _k}:\mathcal Y^n \times {\left[ \left\lfloor {{2^{nM}}} \right\rfloor  \right]} \times \left[N\right]^K \to {\left[ {\left\lceil {{2^{nR}}} \right\rceil } \right]},
\end{equation}
i.e., the reconstructed file by each weak receiver $k \in \left[K_w\right]$ is  
\begin{equation}\label{ReconstructedFileWeak} 
{{\hat W}_{{d_k}}} = {\mu _k}\left( {Y_k^n,{Z_k},\mathbf{d}} \right).
\end{equation}
On the other hand, to serve the users with demand vector $\textbf{d}$, each strong receiver $k \in \left[K_w+1:K\right]$ reconstructs its demand $W_{d_k}$ solely from its channel output $Y_k^n$ through the decoding function      
\begin{equation}\label{DecodingFunctionStrong} 
{\mu _k}:\mathcal Y^n \times \left[N\right]^K \to {\left[ {\left\lceil {{2^{nR}}} \right\rceil } \right]},
\end{equation}
which generates the reconstructed file 
\begin{equation}\label{ReconstructedFileStrong} 
{{\hat W}_{{d_k}}} = {\mu _k}\left( {Y_k^n,\textbf{d}} \right).
\end{equation}

\begin{definition}
An error occurs if ${{\hat W}_{{d_k}}} \ne {W_{{d_k}}}$ for any $k \in \left[ K \right]$, and the probability of error is given by
\begin{equation}\label{ErrorProbability} 
{P_e} \buildrel \Delta \over = \mathop {\max }\limits_{\emph{\textbf{d}} \in \left[ N \right]^K} \Pr \left\{ {\mathop  \bigcup \limits_{k = 1}^K \left\{ {{{\hat W}_{d_k}} \ne {W_{{d_k}}}} \right\}} \right\}.
\end{equation}
\end{definition}

\begin{definition}
A memory-rate pair $(M,R)$ is said to be \textit{achievable}, if for every $\varepsilon > 0$, there exists a large enough $n$, and corresponding caching function \eqref{CachingFunction}, coded delivery function \eqref{DeliveryFunction}, and decoding functions \eqref{DecodingFunctionWeak} and \eqref{DecodingFunctionStrong} at weak and strong receivers, respectively, such that $P_e < \varepsilon$.
\end{definition}

\begin{definition}
For a given cache size $M$ at the weak receivers, the capacity of the network is defined as
\begin{equation}\label{CapcityMemoryTradeoff} 
C \left( M \right) \buildrel \Delta \over = \sup \left\{ R:\mbox{$\left( M,R \right)$ is achievable} \right\}.
\end{equation}
\end{definition}

We would like to note that the capacity of the above caching network remains an open problem even when the delivery channel is an error-free shared bit pipe except for the network with an uncoded cache placement phase \cite{YuMaddahAliAvestimehrExact}. Here, our goal is to identify achievable memory-rate pairs that improve upon the state-of-the-art.

\begin{remark}\label{MotivationSysMod}
The system model considered in this paper is inspired by those studied in \cite{HuangWirelessFadingChannel,ShirinErasureChannelJournal,TimoJointCacheChannelCoding}. We would like to remark that it is reasonable to assume that cache memories are placed at receivers with relatively weaker coverage. Indeed, it is shown in \cite{ShirinErasureChannelJournal} that placing cache memories at the strong receivers, which already have good coverage, results in a lower capacity. This is mainly due to the definition of the capacity in this framework. Note that, the capacity here characterizes the highest rate of the messages to be delivered to all the receivers in the network. Since a symmetry is imposed across the receivers regarding their requests and file delivery, the system performance is determined by the worst receivers. Therefore, the goal of the cache placement should be to improve the performance of the weak receivers to increase the network capacity. Accordingly, equipping weak receivers with cache memories, and exploiting the coding scheme proposed in this paper also benefits the strong receivers.
\end{remark}

Following well-known results from multi-user information theory are included here for completeness as they will be instrumental in deriving our results later in the paper.

\begin{prop}\label{CapRegStandardErasureBCProp}
\cite{UrbankeCapRegPacketBC} The capacity region of a packet erasure BC with $K$ receivers, where file $W_i$ with rate $R_i$ is targeted for receiver $i$ with erasure probability $\delta_i$, for $i=1, ..., K$, is the closure of the set of non-negative rate tuples $\left( R_1, ..., R_K \right)$ that satisfy 
\begin{equation}\label{CapRegStandardErasureBC}
\sum\limits_{i = 1}^K {\frac{{{R_i}}}{{\left( {1 - {\delta _i}} \right)F}}}  \le 1,
\end{equation}
where $F$ denotes the length of the binary channel input.
\end{prop}

Next, we consider the packet erasure BC with side information, and provide an achievable rate pair based on the joint encoding scheme of \cite{TuncelPiggyback}. Here we briefly overview the coding scheme and the proof of achievability, and refer the reader to \cite{TuncelPiggyback} for details. Consider two receivers with erasure probabilities $\delta_1 \ge \delta_2$. Let $W_1$ and $W_2$, distributed uniformly over $\left[ 2^{nR_1} \right]$ and $\left[ 2^{nR_2} \right]$, denote the messages targeted for receivers 1 and 2, respectively. We assume that message $W_2$ is available as side information at receiver 1, the weak receiver. We present a coding scheme and the corresponding achievable rate region based on the joint encoding scheme of \cite{TuncelPiggyback}. For a fixed distribution $P \left( X \right)$, we generate $2^{n \left( R_1+R_2 \right)}$ codewords of length $n$, $x^n \left( w_1,w_2 \right)$, $w_1 \in \left[ 2^{nR_1} \right]$, $w_2 \in \left[ 2^{nR_2} \right]$, where each entry of each codeword is generated independently according to $P \left( X \right)$. The codebook is revealed to the transmitter and the receivers. To transmit particular messages $W_1=w_1$ and $W_2 = w_2$, the codeword $x^n \left( w_1,w_2 \right)$ is transmitted over the BC. In the proposed coding scheme, the good receiver, i.e., receiver 2, decodes both messages; and therefore, it tries to find a unique pair $\left( {\hat w}_1 , {\hat w}_2 \right) \in \left[ 2^{nR_1} \right] \times \left[ 2^{nR_2} \right]$, such that $\left( X^n \left( {\hat w}_1,{\hat w}_2 \right) , y_2^n \right)$ belongs to the jointly typical set defined in \cite{AEGamalNetworkInfTheory}. The probability of decoding error tends to $0$ as $n$ goes to infinity, if
\begin{equation}\label{JointEncodingRateRigionWithoutCache}
R_1 + R_2 \le I \left( X;Y_2 \right).
\end{equation}
The first receiver already knows $W_2$ as side information; therefore, it only needs to decode $W_1$; thus, it looks for a unique index ${\hat w}_1 \in \left[ 2^{nR_1} \right]$ such that $\left( X^n \left( {\hat w}_1,{W}_2 \right) , y_1^n \right)$ belongs to the typical set \cite{AEGamalNetworkInfTheory}. The probability of error tends to $0$ as $n$ goes to infinity, if
\begin{equation}\label{JointEncodingRateRigionWithCache1}
R_1 \le I \left( X;Y_1 \right).
\end{equation}
For the packet erasure BC, both mutual information terms are maximized with a uniform input, and the following conditions are obtained:
\begin{align}\label{CapRegErasureBC}
{R_1} & \le \left( 1-\delta_1 \right) F,\\
{R_1} + {R_2} & \le \left( 1-\delta_2 \right) F,
\end{align}
We can easily generalize this coding scheme to multiple receivers and obtain the achievable rate region stated in the following proposition (also provided in \cite{ShirinErasureChannelJournal}).

\begin{prop}\label{CapRegPiggubackErasureBCProp}
Consider a packet erasure BC with two disjoint sets of receivers $\mathcal{S}_1$ and $\mathcal{S}_2$, where the channels of the receivers in set $\mathcal{S}_i$ have erasure probability $\delta_i$, for $i=1,2$. A common message $W_i$ at rate $R_i$ is to be transmitted to the receivers in set $\mathcal{S}_i$, for $i=1,2$, while message $W_2$ is known to the receivers in set $\mathcal{S}_1$ as side information. With the joint encoding scheme outlined above, rate pairs $\left( R_1, R_2 \right)$ satisfying the following conditions can be achieved
\begin{align}\label{CapRegPiggybackErasureBC}
{R_1} & \le \left( 1-\delta_1 \right) F,\\
{R_1} + {R_2} & \le \left( 1-\delta_2 \right) F,
\end{align}
which is equivalent to
\begin{equation}\label{CapRegPiggybackErasureBCEq}
\max \left\{ {\frac{{{R_1}}}{{\left( 1-\delta_1 \right) F}},\frac{{{R_1} + {R_2}}}{{\left( 1-\delta_2 \right) F}}} \right\} \le 1.
\end{equation}

\end{prop}

For notational convenience, in the rest of the paper we use
\begin{equation}\label{PiggybackNotation}
\mbox{JE}\left( {{{\left( W_1 \right)}_{{\mathcal{S}_1}}},{{\left( W_2 \right)}_{{\mathcal{S}_2}}}} \right)
\end{equation}
to represent the transmission of message $W_1$ to the receivers in set $\mathcal{S}_1$, and message $W_2$ to the receivers in set $\mathcal{S}_2$ using the outlined joint encoding scheme, where ${\mathcal{S}_1} \cap {\mathcal{S}_2} = \emptyset$, and $W_2$ is available at all the receivers in $\mathcal{S}_1$ as side information.

\section{Achievable Rate-Memory Pairs}\label{AchievablePairs}
A coding scheme as well as an information theoretic upper bound on the capacity of the above caching and delivery network are proposed in \cite{ShirinErasureChannelJournal} for the homogeneous scenario. Here, we present a new coding scheme, called the \textit{successive cache-channel coding (SCC)} scheme, for delivery over any packet erasure BC, which is shown to improve upon the one in \cite{ShirinErasureChannelJournal} in the homogeneous scenario. We present the $(M,R)$ pairs achieved by this scheme in Theorem \ref{AchievablePoints} below. The details of the scheme are presented in Section \ref{SCCHeterogeneousHomogeneous}.   

\begin{theorem}\label{AchievablePoints}
Consider cache-aided delivery of $N$ files over a packet erasure BC with $K_w$ weak and $K_s$ strong receivers, where each weak receiver is equipped with a cache of capacity $M$. Memory-rate pairs $\left( {{M_{\left( {p,q} \right)}},{R_{\left( {p,q} \right)}}} \right)$ are achievable for any $p \in \left[ 0:K_w \right]$ and $q \in \left[ p:K_w \right]$, where
\begin{subequations}
\label{AchievableRateMemoryPairsHeter}
\begin{align}\label{AchievableRateMemoryPairsHeter1}
{R_{\left( {p,q} \right)}} \buildrel \Delta \over = & \frac{{F\sum\limits_{i = p}^q { \gamma \left( {p,\boldsymbol{\delta},i} \right)} }}{{\sum\limits_{i = p}^q {\left( {\frac{{ \gamma \left( {p,\boldsymbol{\delta},i} \right)}}{{\binom{K_w}{i} }}\sum\limits_{j = 1}^{{K_w} - i} {\frac{{\binom{K_w-j}{i} }}{{1 - {\delta _j}}}} } \right)}  + \sum\limits_{j = {K_w} + 1}^{{K}} {\frac{1}{{1 - {\delta _j}}}} }},\\
{{M}_{\left( {p,q} \right)}} \buildrel \Delta \over = & \frac{N{\sum\limits_{i = p}^q {i \gamma \left( {p,\boldsymbol{\delta},i} \right)} }}{K_w{\sum\limits_{i = p}^q { \gamma \left( {p,\boldsymbol{\delta},i} \right)} }}{{ R}_{\left( {p,q} \right)}},
\label{AchievableRateMemoryPairsHeter2}
\end{align}
with $\gamma(p,\boldsymbol{\delta},i)$ defined as follows:
\begin{align}\label{gammapiHeter}
\gamma(p,\boldsymbol{\delta},i) \buildrel \Delta \over = \frac{\binom{K_w}{i}}{{\binom{K_w}{p}{K_s}^{i - p}}}\prod\limits_{j = p}^{i - 1} \left( \frac{{{K_s}}}{{\left( {1 - {\delta _{{K_w} - j}}} \right)\sum\limits_{l = {K_w} + 1}^K {\frac{1}{{1 - {\delta _l}}}} }} - 1 \right), \quad \mbox{for $i=p,...,q$}.
\end{align}
\end{subequations}
The upper convex hull of these $\left( {{K_w} + 1} \right)\left( {{K_w} + 2} \right)/2$ memory-rate pairs can also be achieved through memory-sharing. 
\end{theorem}

\begin{corollary}\label{RemarkAchievableMemoryRtaePairHomogeneous}
For the homogeneous scenario, the achievable memory-rate pairs $\left( {{M_{\left( {p,q} \right)}},{R_{\left( {p,q} \right)}}} \right)$, for any $p \in \left[ 0:K_w \right]$ and $q \in \left[ p:K_w \right]$, are simplified as follows:
\begin{subequations}
\label{AchievableRateMemoryPairsHomog}
\begin{align}\label{AchievableRateMemoryPairsHomog1}
{R_{\left( {p,q} \right)}} = & \frac{{F\sum\limits_{i = p}^q {{\gamma \left( {p,{\boldsymbol{\delta}}_{ws},i} \right)}} }}{{\frac{1}{{1 - {\delta _w}}}\sum\limits_{i = p}^q {\left( {\frac{{{K_w} - i}}{{i + 1}}\gamma \left( {p,{\boldsymbol{\delta}}_{ws},i} \right)} \right)}  + \frac{{{K_s}}}{{1 - {\delta _s}}}}},\\
{{M}_{\left( {p,q} \right)}} = & \frac{N{\sum\limits_{i = p}^q {i \gamma \left( {p,{\boldsymbol{\delta}}_{ws},i} \right)} }}{K_w{\sum\limits_{i = p}^q {\gamma \left( {p,{\boldsymbol{\delta}}_{ws},i} \right)} }}{{R}_{\left( {p,q} \right)}},
\label{AchievableRateMemoryPairsHomog2}
\end{align}
where
\begin{align}\label{gammapiHomog}
\gamma \left(p,{\boldsymbol{\delta}}_{ws},i \right) = {\frac{\binom{K_w}{i}}{{\binom{K_w}{p}{K_s}^{i - p}}}{{\left( {\frac{{1 - {\delta _s}}}{{1 - {\delta _w}}} - 1} \right)}^{i - p}}}, \quad \mbox{for $i=p,...,q$}.
\end{align}
\end{subequations}
\end{corollary}

\begin{remark}\label{RemarkInsightOnGamma}
As we will explain in Section \ref{SCCHeterogeneousHomogeneous}, to achieve the rate-memory pair $\left( {{{ M}_{\left( {p,q} \right)}},{{ R}_{\left( {p,q} \right)}}} \right)$ with the proposed SCC scheme, for $p \in \left[ 0:K_w \right]$ and $q \in \left[ p:K_w \right]$, each file is divided into $p-q+1$ non-overlapping subfiles. Here we remark that $\gamma \left(p,\boldsymbol{\delta},i \right)/{\sum\limits_{j = p}^q { { \gamma \left( {p,\boldsymbol{\delta},j} \right)} } }$, for $i \in [p:q]$, indicates the fraction of the total rate $R$ allocated to the $(i-p+1)$-th of these subfiles. Thus, a proportioned rate allocation is performed according to the coefficients $\gamma \left(p,\boldsymbol{\delta},p\right), \gamma \left(p,\boldsymbol{\delta},p+1\right), ..., \gamma \left(p,\boldsymbol{\delta},q\right)$.  
\end{remark}

\begin{figure}[!t]
\centering
\includegraphics[scale=0.5]{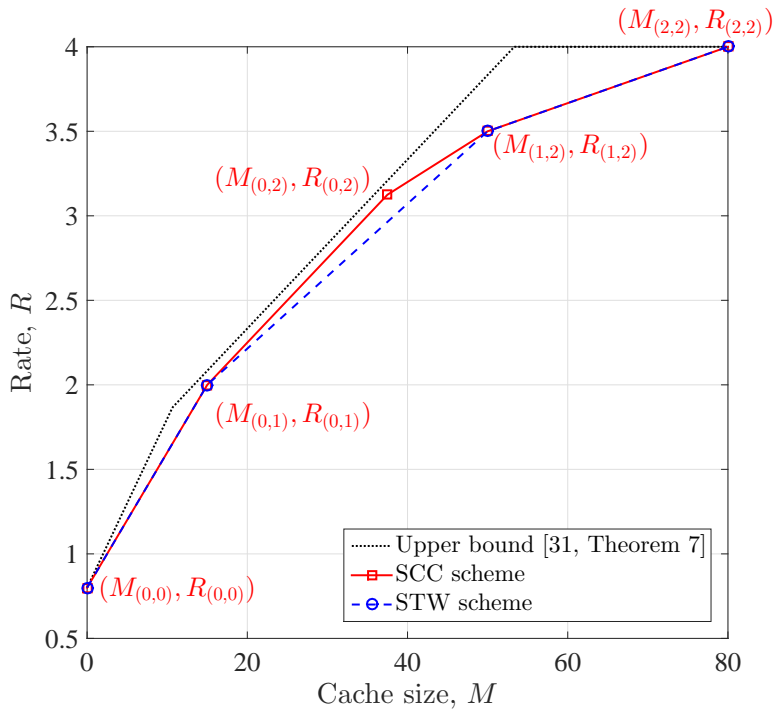}
\caption{Lower and upper bounds on the capacity for the homogeneous scenario with $K_w=2$, $K_s=2$, $N=20$, $F=10$, $\delta_s = 0.2$, and $\delta_w = 0.8$.} 
\label{Ks2_Kw2}
\end{figure}

Next, we compare the achievable rate of the SCC scheme for the homogeneous scenario with the scheme of \cite{ShirinErasureChannelJournal}, which we will refer to as the STW scheme. In Fig. \ref{Ks2_Kw2}, the achievable memory-rate trade-off of the SCC scheme is compared with the STW scheme when $K_w=2$, $K_s=2$, $N=20$, $F=10$, $\delta_s = 0.2$, and $\delta_w = 0.8$. The upper bound on the capacity of the cache-aided packet erasure BC derived in \cite[Theorem 7]{ShirinErasureChannelJournal} is also included. The SCC scheme outperforms the STW scheme due to the improved achievable memory-rate pair $\left( M_{(0,2)}, R_{(0,2)}\right)$, which is not achievable by the STW scheme. This improvement can be extended to a wider range of cache sizes through memory-sharing, also reducing the gap to the upper bound. 

\begin{figure}[!t]
\centering
\includegraphics[scale=0.5]{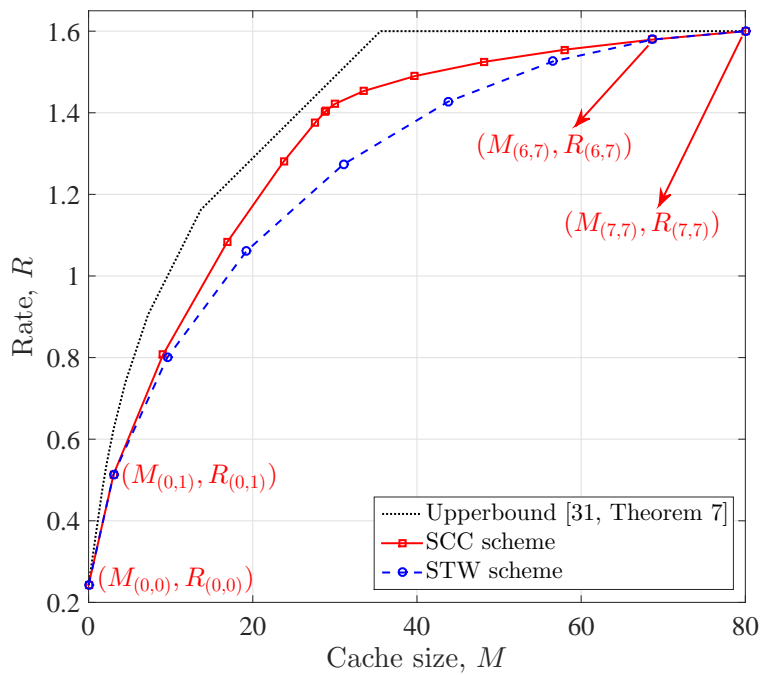}
\caption{Lower and upper bounds on the capacity for the homogeneous scenario with $K_w=7$, $K_s=10$, $N=50$, $F=20$, $\delta_s = 0.2$, and $\delta_w = 0.9$.} 
\label{Ks15_Kw10}
\end{figure}

In Fig. \ref{Ks15_Kw10}, we plot the achievable rates for both schemes in the homogeneous scenario with $K_w=7$, $K_s=10$, $N=50$, $F=20$, $\delta_s = 0.2$, and $\delta_w = 0.9$. The upper bound on the capacity derived in \cite[Theorem 7]{ShirinErasureChannelJournal} is also included. Observe that, for relatively small cache sizes, where the best memory-rate trade-off is achieved by time-sharing between $\left( M_{(0,0)}, R_{(0,0)}\right)$ and $\left( M_{(0,1)}, R_{(0,1)}\right)$, and for relatively large cache sizes, where the best memory-rate trade-off is achieved by time-sharing between $\left( M_{(6,7)}, R_{(6,7)}\right)$ and $\left( M_{(7,7)}, R_{(7,7)}\right)$, both schemes achieve the same rate; however, the proposed SCC scheme achieves a higher rate than STW scheme for all other intermediate cache sizes, and reduces the gap to the upper bound. For a cache capacity of $M=30$, the SCC scheme provides approximately $15 \%$ increase in the achievable rate compared to the STW scheme.

\begin{figure}[!t]
\centering
\includegraphics[scale=0.5]{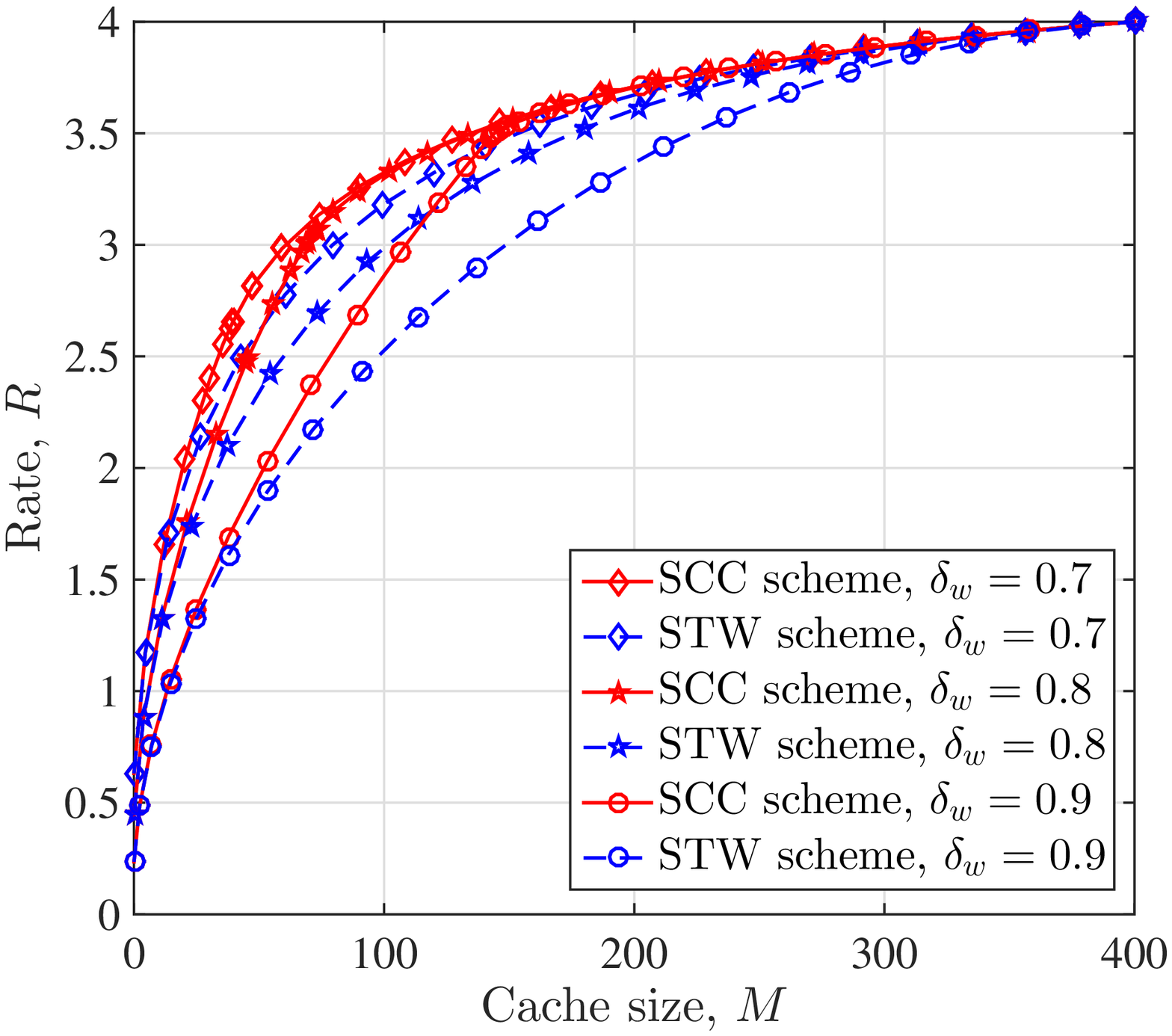}
\caption{Lower and upper bounds on the capacity for the homogeneous scenario with $K_w=20$, $K_s=10$, $N=100$, $F=50$, $\delta_s = 0.2$, and different values for $\delta_w$ given by $\delta_w =0.7, 0.8, 0.9$.} 
\label{Ks10_Kw20_delta_w}
\end{figure}

In Fig. \ref{Ks10_Kw20_delta_w}, the achievable rates of the SCC and STW schemes in the homogeneous scenario are compared for different values of $\delta_w$. System parameters considered in this comparison are $K_w=20$, $K_s=10$, $N=100$, $F=50$, $\delta_s=0.2$, and $\delta_w=0.7, 0.8, 0.9$. Observe that, unlike the STW scheme, the performance of the SCC scheme does not deteriorate notably for intermediate and relatively high cache capacities when $\delta_w$ increases, i.e., having worse channel qualities for the weak receivers. This is because the SCC scheme successfully exploits the available cache capacities, and there is little to lose from increasing $\delta_w$ when $M$ is sufficiantly large. Moreover, the superiority of the SCC scheme over the STW scheme is more pronounced for higher values of $\delta_w$, in which case, exploiting the cache memories available at the weak receivers more effectively through SCC becomes more important.

For the heterogeneous scenario, the capacity of the network under consideration is upper bounded by \cite{ShirinErasureChannelJournal}
\begin{equation}\label{UpperboundHeter}
\mathop {\min }\limits_{\mathcal{S} \subset \left[ K \right]} \left\{ {F{{\left( {\sum\limits_{k \in \mathcal{S}} {\frac{1}{{1 - {\delta _k}}}} } \right)}^{ - 1}} + \frac{M}{N}\left| {\mathcal{S} \cap \left[ {{K_w}} \right]} \right|} \right\}.
\end{equation}



\begin{figure}[!t]
\centering
\includegraphics[scale=0.5]{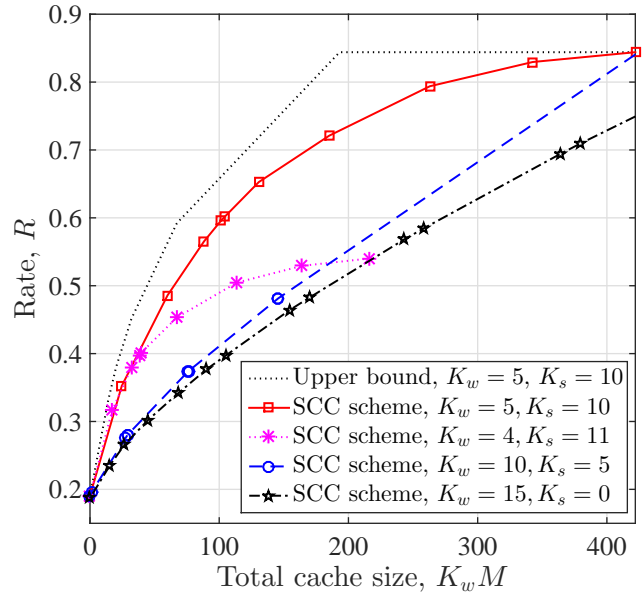}
\caption{Lower and upper bounds on the capacity for the heterogeneous scenario with $K=15$, $N=100$, $F=10$, $\delta_k = 0.9 - 0.01k$, for $k=1, ..., 5$, and $\delta_{l} = 0.2 - 0.01l$, for $l=6, ..., 15$ with variable $K_w$ and $K_s$.} 
\label{Variable_Ks_Kw}
\end{figure}

In Fig. \ref{Variable_Ks_Kw}, the effect of $K_w$ is considered for the heterogeneous scenario with the system parameters $K=15$, $N=100$, $F=10$, $\delta_k = 0.9 - 0.01k$, for $k=1, ..., 5$, and $\delta_{l} = 0.2 - 0.01l$, for $l=6, ..., 15$. In this figure, the achievable rates are plotted with respect to the total cache capacity of $K_wM$ for four different values for the number of weak receivers in the system, $K_w=4,5,10,15$. Note that the erasure probabilities are set such that the first $5$ receivers have significantly worse channels than the remaining $10$ receivers. Note also that the parameter $K_w$ determines which receivers are provided with cache memories. As it can be seen, the setting with $K_w=5$ achieves significantly higher rates over a wide range of total cache capacities compared to the other settings under consideration. If receiver $5$, which has a relatively bad channel quality, is not provided with any cache memory, and only the first 4 receivers are equipped with cache memories, i.e., $K_w=4$, the performance degrades significantly except for very small values of total cache size. This is because the first five receivers have much worse channel qualities, and the performance depends critically on the caches provided to all these five weak receivers. On the other hand, equipping receivers with relatively good channel qualities with cache memories deteriorates the performance of the system in terms of the achievable rate. Note that this is because the total available cache capacity is allocated across a larger number of receivers. This result confirms that it is more beneficial to allocate cache memories to the receivers with relatively worse channel qualities. The upper bound on the achievable rate for the setting with $K_w=5$ and $K_s=10$ is also included in this figure. We observe that the gap between the upper bound and the achievable rate for the same setting is relatively small for a wide range of cache sizes.

\section{The Successive Cache-Channel Coding (SCC) Scheme}\label{SCCHeterogeneousHomogeneous}

Before presenting the SCC scheme for the general heterogeneous scenario, in which we allow the weak and strong receivers to have distinct erasure probabilities, the main ideas behind this scheme are illustrated on an example in the simplified homogeneous scenario. 

For notational convenience, the $i$-element subsets of set $[K_w]$ are enumerated by $\mathcal{S}_{1}^{(i)}, \mathcal{S}_{2}^{(i)},...,$ $ \mathcal{S}_{\binom{K_w}{i}}^{(i)}$, i.e.,
\begin{equation}\label{SubsetsofKw}
\mathcal{S}_{j}^{(i)} \subset [K_w] \mbox{ and } \left| \mathcal{S}_{j}^{(i)} \right|=i, \quad \mbox{for $i \in [0:K_w]$, and $j=1, ..., \binom{K_w}{i}$}.
\end{equation}

\theoremstyle{definition}
\newtheorem{exmp}{Example}

\begin{exmp}\label{exampleGeneralCase}
Consider the cache-aided packet erasure homogeneous BC depicted in Fig. \ref{System_Model} with $K_w=3$ weak and $K_s=2$ strong receivers. In the following, we investigate the achievable memory-rate pair $\left( {{M_{\left( {0,2} \right)}},{R_{\left( {0,2} \right)}}} \right)$, which corresponds to the memory-rate pair in \eqref{AchievableRateMemoryPairsHomog} for $p=0$ and $q=2$. Each file $W_f$, $f \in \left[ N \right]$, is divided into three subfiles $W_f^{(0)}$, $W_f^{(1)}$ and $W_f^{(2)}$, where subfile $W_f^{(i)}$ has a rate of $R^{(i)}$, for $i=0,1,2$, given by
\begin{equation}\label{EachRiExmp}
R^{(i)} \buildrel \Delta \over = \frac{\gamma \left( 0,{\boldsymbol{\delta}}_{ws},i \right)}{\sum\limits_{j = 0}^2 \gamma \left( 0,{\boldsymbol{\delta}}_{ws},j \right) }R,
\end{equation}
where $\gamma \left(0,{\boldsymbol{\delta}}_{ws},i\right)$ is as defined in \eqref{gammapiHomog}. We have $\sum\limits_{i = 0}^2 {{R^{\left( i \right)}}}  = R$.

\textbf{Placement phase:} In the placement phase, subfiles ${W_1^{\left( i \right)},...,W_N^{\left( i \right)}}$ are placed in the caches of $K_w=3$ weak receivers using the procedure in \cite[Algorithm 1]{MaddahAliCentralized}, specified for a cache capacity of $iN/K_w$, for $i=0,1,2$. In this cache placement procedure, each subfile $W_f^{(i)}$ is first divided into $\binom{3}{i}$ non-overlapping pieces, each at a rate of $R^{(i)} / \binom{3}{i}$. 
\begin{align}\label{DivideEachSubfileExample}
W_f^{\left( i \right)} = &\left( {W_{f,\mathcal{S}^{(i)}_1}^{\left( i \right)}, W_{f,\mathcal{S}^{(i)}_2}^{\left( i \right)}, ..., W_{f,\mathcal{S}^{(i)}_{\binom{3}{i}}}^{\left( i \right)}} \right),  \forall f \in \left[ N \right], \forall i \in \left[ 0:2 \right],
\end{align}
For the example under consideration, we have, $\forall f \in \left[ N \right]$,    
\begin{subequations}
\label{PiecesExmp}
\begin{align}\label{PiecesExmp1}
W_f^{\left( 0 \right)} & = \left( {W_{f,\emptyset}^{\left( 0 \right)}} \right) ,\\
W_f^{\left( 1 \right)} & = \left( {W_{f,\left\{ 1 \right\}}^{\left( 1 \right)},W_{f,\left\{ 2 \right\}}^{\left( 1 \right)},W_{f,\left\{ 3 \right\}}^{\left( 1 \right)}}   \right), 
\label{PiecesExmp2}\\
W_f^{\left( 2 \right)} & = \left( W_{f,\left\{ {1,2} \right\}}^{\left( 2 \right)}, W_{f,\left\{ {1,3} \right\}}^{\left( 2 \right)}, W_{f,\left\{ {2,3} \right\}}^{\left( 2 \right)} \right).
\label{PiecesExmp3}
\end{align}
\end{subequations}
The piece ${W_{f,\mathcal{S}^{(i)}_l}^{\left( i \right)}}$ is placed in the cache of each receiver $k \in \mathcal{S}^{(i)}_l$, for $l=1, ..., \binom{3}{i}$. Therefore, the cache contents of the weak receivers after the placement phase are as follows:
\begin{subequations}
\label{CacheContentsExmp}
\begin{align}\label{CacheContentsExmp1}
{Z_1} & = \bigcup\limits_{f \in \left[ N \right]} {\left( {W_{f,\left\{ 1 \right\}}^{\left( 1 \right)},W_{f,\left\{ {1,2} \right\}}^{\left( 2 \right)},W_{f,\left\{ {1,3} \right\}}^{\left( 2 \right)}} \right)},\\
{Z_2} & = \bigcup\limits_{f \in \left[ N \right]} {\left( {W_{f,\left\{ 2 \right\}}^{\left( 1 \right)},W_{f,\left\{ {1,2} \right\}}^{\left( 2 \right)},W_{f,\left\{ {2,3} \right\}}^{\left( 2 \right)}} \right)} ,\label{CacheContentsExmp2}\\
{Z_3} & = \bigcup\limits_{f \in \left[ N \right]} {\left( {W_{f,\left\{ 3 \right\}}^{\left( 1 \right)},W_{f,\left\{ {1,3} \right\}}^{\left( 2 \right)},W_{f,\left\{ {2,3} \right\}}^{\left( 2 \right)}} \right)},
\label{CacheContentsExmp3}
\end{align}
\end{subequations}
where the required cache capacity for each weak receiver is:
\begin{align}\label{CacheCapacityExmp}
M = \left( {\frac{{{R^{\left( 1 \right)}}}}{3} + \frac{{{2R^{\left( 2 \right)}}}}{3}} \right)N= \frac{{ {\gamma \left( {0,{\boldsymbol{\delta}}_{ws},1} \right) + 2\gamma \left( {0,{\boldsymbol{\delta}}_{ws},2} \right)} }}{{3\sum\limits_{j = 0}^2 { {\gamma \left( {0,{\boldsymbol{\delta}}_{ws},j} \right)}} }}NR.
\end{align}

\textbf{Delivery phase:} The server tries to satisfy all the demands in the delivery phase by sending four distinct messages in an orthogonal fashion, i.e., by time division multiplexing, where the codewords corresponding to the $i$-th message, $i=1, ..., 4$, are of length $\beta_i n$ channel uses, such that $\sum\limits_{i = 1}^4 {{\beta _i}}  = 1$. The contents delivered with each message are illustrated in Table \ref{TableDeliveryExample1}.

The first message is targeted only for the weak receivers, and its goal is to deliver the missing subfiles of file $W_{{d_k}}^{\left( 2 \right)}$ to receiver $k$, $k=1, 2, 3$, that is, having received this message, each weak receiver should be able to decode the third subfile of its desired file. Exploiting the delivery phase of \cite[Algorithm 1]{MaddahAliCentralized} for cache capacity $2N/K_w$, the coded content with message $1$ in Table \ref{TableDeliveryExample1} is delivered to the weak receivers $\{1,2,3\}$. Having received message 1 given in Table \ref{TableDeliveryExample1}, receiver $k$ can recover its missing piece $W_{{d_k},\left[ 3 \right]\backslash \{k\}}^{\left( 2 \right)}$ of subfile $W_{{d_k}}^{\left( 2 \right)}$ using its cache contents $Z_k$. Thus, together with its cache content, receiver $k$ can recover subfile $W_{{d_k}}^{\left( 2 \right)}$, for $k=1, 2,3$.

\begin{table}[!t]
\caption{Contents sent with messages $1$ to $4$ in the delivery phase of Example \ref{exampleGeneralCase}.}
\centering
\begin{tabular}{ ||c||c|c|| }
\hline
Message 1 & \multicolumn{2}{c||}{$\left( W_{{d_1},\left\{ {2,3} \right\}}^{\left( 2 \right)} \oplus W_{{d_2},\left\{ {1,3} \right\}}^{\left( 2 \right)} \oplus W_{{d_3},\left\{ {1,2} \right\}}^{\left( 2 \right)} \right)$ to receivers $1,2,3$}\\
\hline \hline
\multirow{5.1}{*}{Message 2} & Sub-message 1 & $\mbox{JE} \left( {\left( {W_{{d_1},\left\{ 2 \right\}}^{\left( 1 \right)} \oplus W_{{d_2},\left\{ 1 \right\}}^{\left( 1 \right)}} \right)}_{\left\{ {1,2} \right\}},  {\left( {W_{{d_4},\left\{ {1,2} \right\}}^{\left( 2 \right)},W_{{d_5},\left\{ {1,2} \right\}}^{\left( 2 \right)}} \right)}_{\left\{ {4,5} \right\}} \right)$\\
\cline{2-3}
& Sub-message 2 & $\mbox{JE} \left( {\left( {W_{{d_1},\left\{ 3 \right\}}^{\left( 1 \right)} \oplus W_{{d_3},\left\{ 1 \right\}}^{\left( 1 \right)}} \right)}_{\left\{ {1,3} \right\}}, {\left( {W_{{d_4},\left\{ {1,3} \right\}}^{\left( 2 \right)},W_{{d_5},\left\{ {1,3} \right\}}^{\left( 2 \right)}} \right)}_{\left\{ {4,5} \right\}} \right)$\\
\cline{2-3}
& Sub-message 3 & $\mbox{JE} \left( {\left( {W_{{d_2},\left\{ 3 \right\}}^{\left( 1 \right)} \oplus W_{{d_3},\left\{ 2 \right\}}^{\left( 1 \right)}} \right)}_{\left\{ {2,3} \right\}}, {\left( {W_{{d_4},\left\{ {2,3} \right\}}^{\left( 2 \right)},W_{{d_5},\left\{ {2,3} \right\}}^{\left( 2 \right)}} \right)}_{\left\{ {4,5} \right\}} \right)$\\
\hline \hline
\multirow{5.1}{*}{Message 3} & Sub-message 1 & $\mbox{JE}\left( {{{\left( {W_{{d_1},\emptyset}^{\left( 0 \right)}} \right)}_{\left\{ {1} \right\}}},{{\left( {W_{{d_4},\left\{ {1} \right\}}^{\left( 1 \right)},W_{{d_5},\left\{ {1} \right\}}^{\left( 1 \right)}} \right)}_{\left\{ {4,5} \right\}}}} \right)$\\
\cline{2-3}
& Sub-message 2 & $\mbox{JE}\left( {{{\left( {W_{{d_2},\emptyset}^{\left( 0 \right)}} \right)}_{\left\{ {2} \right\}}},{{\left( {W_{{d_4},\left\{ {2} \right\}}^{\left( 1 \right)},W_{{d_5},\left\{ {2} \right\}}^{\left( 1 \right)}} \right)}_{\left\{ {4,5} \right\}}}} \right)$\\
\cline{2-3}
& Sub-message 3 & $\mbox{JE}\left( {{{\left( {W_{{d_3},\emptyset}^{\left( 0 \right)}} \right)}_{\left\{ {3} \right\}}},{{\left( {W_{{d_4},\left\{ {3} \right\}}^{\left( 1 \right)},W_{{d_5},\left\{ {3} \right\}}^{\left( 1 \right)}} \right)}_{\left\{ {4,5} \right\}}}} \right)$\\
\hline \hline
Message 4 & \multicolumn{2}{|c||}{$\left( {W_{{d_4},\emptyset}^{\left( 0 \right)}} \right)$ to receiver $4$, and $\left( {W_{{d_5},\emptyset}^{\left( 0 \right)}} \right)$ to receiver $5$}\\
\hline
\end{tabular}
\label{TableDeliveryExample1}
\end{table}

Through the second message of the delivery phase, the server simultaneously delivers subfile $W^{(2)}_{d_l}$ to strong receiver $l$, $l=4,5$, and the missing bits of subfile $W^{(1)}_{d_k}$ to weak receiver $k$, $k=1,2,3$. The content targeted to the weak receivers is delivered by using the delivery phase of \cite[Algorithm 1]{MaddahAliCentralized} for the cache capacity of $N/K_w$; that is, the contents 
\begin{equation}\label{DeliveredCodedContentsWeakPart2Exmp}
\left\{ {W_{{d_1},\left\{ 2 \right\}}^{\left( 1 \right)} \oplus W_{{d_2},\left\{ 1 \right\}}^{\left( 1 \right)},W_{{d_1},\left\{ 3 \right\}}^{\left( 1 \right)} \oplus W_{{d_3},\left\{ 1 \right\}}^{\left( 1 \right)},W_{{d_2},\left\{ 3 \right\}}^{\left( 1 \right)} \oplus W_{{d_3},\left\{ 2 \right\}}^{\left( 1 \right)}} \right\}
\end{equation}
are transmitted to the weak receivers. Therefore, the goal is to deliver $W_{d_l}^{(2)}$ to strong receiver $l$, $l=4,5$, while delivering the contents in \eqref{DeliveredCodedContentsWeakPart2Exmp} to the weak receivers in parallel. The transmission is performed by sending three sub-messages, transmitted over orthogonal time periods. With the first sub-message of message 2 given in Table \ref{TableDeliveryExample1}, receivers 1 and 2 receive $W_{{d_1},\left\{ 2 \right\}}^{\left( 1 \right)} \oplus W_{{d_2},\left\{ 1 \right\}}^{\left( 1 \right)}$ since they both have $W_{{d_4},\left\{ {1,2} \right\}}^{\left( 2 \right)}$ and $W_{{d_5},\left\{ {1,2} \right\}}^{\left( 2 \right)}$ in their caches as side information. Accordingly, receiver 1 and receiver 2 can recover $W_{{d_1},\left\{ 2 \right\}}^{\left( 1 \right)}$ and $W_{{d_2},\left\{ 1 \right\}}^{\left( 1 \right)}$, respectively. On the other hand, with the joint encoding scheme, $W_{{d_4},\left\{ {1,2} \right\}}^{\left( 2 \right)}$ and $W_{{d_5},\left\{ {1,2} \right\}}^{\left( 2 \right)}$ are directly delivered to receiver 4 and receiver 5, respectively. With the second sub-message of message 2 in Table \ref{TableDeliveryExample1}, $W_{{d_4},\left\{ {1,3} \right\}}^{\left( 2 \right)}$ and $W_{{d_5},\left\{ {1,3} \right\}}^{\left( 2 \right)}$, which are available in the caches of receivers $1$ and $3$ as side information, are delivered to receivers $4$ and $5$, while $W_{{d_1},\left\{ 3 \right\}}^{\left( 1 \right)} \oplus W_{{d_3},\left\{ 1 \right\}}^{\left( 1 \right)}$ is delivered to receivers $1$ and $3$. By receiving sub-message 2, receiver 1 and receiver 3 can obtain $W_{{d_1},\left\{ 3 \right\}}^{\left( 1 \right)}$ and $W_{{d_3},\left\{ 1 \right\}}^{\left( 1 \right)}$, respectively. Finally, sub-message 3 of message 2 aims to deliver $W_{{d_4},\left\{ {2,3} \right\}}^{\left( 2 \right)}$ and $W_{{d_5},\left\{ {2,3} \right\}}^{\left( 2 \right)}$, which are in the cache of receivers 2 and 3, to receivers $4$ and $5$, respectively, and $W_{{d_2},\left\{ 3 \right\}}^{\left( 1 \right)} \oplus W_{{d_3},\left\{ 2 \right\}}^{\left( 1 \right)}$ to receivers $2$ and $3$ by the joint encoding scheme. Having received coded content $W_{{d_2},\left\{ 3 \right\}}^{\left( 1 \right)} \oplus W_{{d_3},\left\{ 2 \right\}}^{\left( 1 \right)}$, receiver 2 and receiver 3 can recover $W_{{d_2},\left\{ 3 \right\}}^{\left( 1 \right)}$ and $W_{{d_3},\left\{ 2 \right\}}^{\left( 1 \right)}$, respectively. Thus, having received message 2, each weak receiver $k$, $k=1,2,3$, can recover all the missing bits of subfile $W_{d_k}^{(1)}$ of its request, while each strong receiver $l$, $l=4,5$, can obtain subfile $W_{d_l}^{(2)}$ of its request.

The third message of the delivery phase is designed to deliver $W_{d_l}^{(1)}$ to strong receiver $l$, $l=4,5$, and ${W_{{d_k}}^{\left( 0 \right)}}$ is delivered to weak receiver $k$, $k=1,2,3$. Third message is also divided into three sub-messages, transmitted over orthogonal time periods. With sub-message $k$, given in Table \ref{TableDeliveryExample1}, $W_{{d_4},\left\{ {k} \right\}}^{\left( 1 \right)}$ and $W_{{d_5},\left\{ {k} \right\}}^{\left( 1 \right)}$, both of which are available locally at receiver $k$ as side information, $k=1,2,3$, are delivered to receivers $4$ and $5$, respectively, while ${W_{{d_k},\emptyset}^{\left( 0 \right)}}$ is delivered to receiver $k$. Therefore, with the third message in Table \ref{TableDeliveryExample1}, each weak receiver $k$, $k=1,2,3$, can obtain ${W_{{d_k},\emptyset}^{\left( 0 \right)}}$, while each strong receiver $l$, $l=4,5$, can recover $W_{d_l}^{(1)}$. Thus, after receiving message 3 in Table \ref{TableDeliveryExample1}, the demands of the weak receivers are fully satisfied.

The last and fourth message of the delivery phase is generated only for the strong receivers with the goal of delivering them the missing bits of their demands, in particular, subfile ${W_{{d_l},\emptyset}^{\left( 0 \right)}}$ is delivered to each strong receiver $l$, $l=4,5$.

Observe that message 1 in Table \ref{TableDeliveryExample1} has a rate of ${R^{\left( 2 \right)}}/3$. The capacity region of the standard packet erasure BC presented in Proposition \ref{CapRegStandardErasureBCProp} suggests that all the weak receivers can decode message 1, for $n$ large enough, if
\begin{equation}\label{AchievableDeliveryRatePart1Exmp}
\frac{{{R^{\left( 2 \right)}}}}{{3\left( {1 - {\delta _w}} \right)F}} \le {\beta _1}.
\end{equation}

From Table \ref{TableDeliveryExample1}, with each sub-message of the second message, messages of rate $2R^{(2)}/3$, available at the weak receivers as side information, are delivered to the strong receivers; while, simultaneously, a common message at rate $R^{(1)}/3$ is transmitted to the weak receivers. Overall, $\left( W_{d_4}^{(2)}, W_{d_5}^{(2)} \right)$ and the contents in \eqref{DeliveredCodedContentsWeakPart2Exmp} with a total rate of $2R^{(2)}$ and $R^{(1)}$ are delivered to the strong and weak receivers, respectively, through three different sub-messages by using the joint encoding scheme of \cite{TuncelPiggyback} that exploits the side information at the weak receivers. Using the achievable rate region of the joint encoding scheme for the packet erasure channels stated in Proposition \ref{CapRegPiggubackErasureBCProp}, $\left( W_{d_4}^{(2)}, W_{d_5}^{(2)} \right)$ and the contents in \eqref{DeliveredCodedContentsWeakPart2Exmp} can be simultaneously decoded by the strong and weak receivers, respectively, for $n$ large enough, if
\begin{equation}\label{CapacityRegionPiggybackPart2Exmp}
\max \left\{ {\frac{{{R^{\left( 1 \right)}}}}{{\left( {1 - {\delta _w}} \right)F}},\frac{R^{\left( 1 \right)}+{2{R^{\left( 2 \right)}}}}{{\left( {1 - {\delta _s}} \right)F}}} \right\} \le {\beta _2}.
\end{equation}
From the expressions for $R^{(1)}$ and $R^{(2)}$ in \eqref{EachRiExmp}, it can be verified that the two terms in the maximization in \eqref{CapacityRegionPiggybackPart2Exmp} are equal for the setting under consideration. Thus, the condition in \eqref{CapacityRegionPiggybackPart2Exmp} can be simplified as
\begin{equation}\label{CapacityRegionPiggybackFinalPart2Exmp}
\frac{{{R^{\left( 1 \right)}}}}{{\left( {1 - {\delta _w}} \right)F}} \le {\beta _2}.
\end{equation}

According to Table \ref{TableDeliveryExample1}, with each sub-message of message 3, a message at rate $R^{(0)}$ is targeted for the weak receivers, while message at rate $2R^{(1)}/3$, available locally at the weak receivers, is aimed for the strong receivers. Therefore, through joint encoding scheme over three periods, messages with a total rate of $3R^{(0)}$ are delivered to the weak receivers, while the strong receivers receive a total rate of $2R^{(1)}$. According to Proposition \ref{CapRegPiggubackErasureBCProp}, all the weak and strong receivers can decode their messages, for $n$ large enough, if
\begin{equation}\label{CapacityRegionPiggybackPart3Exmp}
\max \left\{ {\frac{{{3R^{\left( 0 \right)}}}}{{\left( {1 - {\delta _w}} \right)F}},\frac{3R^{\left( 0 \right)}+{2{R^{\left( 1 \right)}}}}{{\left( {1 - {\delta _s}} \right)F}}} \right\} \le {\beta _3}.
\end{equation}
Again, from the expressions of $R^{(0)}$ and $R^{(1)}$ in \eqref{EachRiExmp}, it can be verified that, when $K_w=3$ and $K_s=2$, \eqref{CapacityRegionPiggybackPart3Exmp} can be simplified as
\begin{equation}\label{CapacityRegionPiggybackFinalPart3Exmp}
\frac{{3{R^{\left( 0 \right)}}}}{{\left( {1 - {\delta _w}} \right)F}} \le {\beta _3}.
\end{equation}

From the capacity region of the standard erasure BC in Proposition \ref{CapRegStandardErasureBCProp}, each receiver $l$, $l=4,5$, can decode ${W_{{d_l},\emptyset}^{\left( 0 \right)}}$, delivered with message 4, successfully for $n$ sufficiently large, if
\begin{equation}\label{CapacityRegionPiggybackPart4Exmp}
\frac{{2{R^{\left( 0 \right)}}}}{{\left( {1 - {\delta _s}} \right)F}} \le {\beta _4}.
\end{equation}
Combining \eqref{AchievableDeliveryRatePart1Exmp}, \eqref{CapacityRegionPiggybackFinalPart2Exmp}, \eqref{CapacityRegionPiggybackFinalPart3Exmp}, \eqref{CapacityRegionPiggybackPart4Exmp}, and the fact that $\sum\limits_{i = 1}^4 {{\beta _i}}  = 1$, we have the condition    
\begin{equation}\label{AchievableRateExmp}
\frac{{{R^{\left( 2 \right)}}}}{{3\left( {1 - {\delta _w}} \right)F}} + \frac{{{R^{\left( 1 \right)}}}}{{\left( {1 - {\delta _w}} \right)F}} + \frac{{3{R^{\left( 0 \right)}}}}{{\left( {1 - {\delta _w}} \right)F}} + \frac{{2{R^{\left( 0 \right)}}}}{{\left( {1 - {\delta _s}} \right)F}} \le 1.
\end{equation}
By replacing $R^{(i)}$ with the expressions from \eqref{EachRiExmp}, for $i=0,1,2$, and using the fact that $\gamma \left(0,{\boldsymbol{\delta}}_{ws},0 \right)=1$, \eqref{AchievableRateExmp} is reduced to
\begin{equation}\label{AchievableRateFinalExmp}
R \le \frac{{\sum\limits_{j = 0}^2 { {\gamma \left( {0,{\boldsymbol{\delta}}_{ws},j} \right)}} F}}{{\frac{1}{{1 - {\delta _w}}}\left( {3 + \gamma \left( {0,{\boldsymbol{\delta}}_{ws},1} \right) + \frac{1}{3}\gamma \left( {0,{\boldsymbol{\delta}}_{ws},2} \right)} \right) + \frac{2}{{1 - {\delta _s}}}}}.
\end{equation}
Observe that, the term on the right hand side of inequality \eqref{AchievableRateFinalExmp} is $R_{(0,2)}$, which is given by \eqref{AchievableRateMemoryPairsHomog1}. The cache size of each weak receiver exploited by our coding scheme is given by \eqref{CacheCapacityExmp}, and we have $M=M_{(0,2)}$, where $M_{(0,2)}$ is defined as in \eqref{AchievableRateMemoryPairsHomog2}. Thus, the memory-rate pair $\left( M_{\left( 0,2 \right)}, \right.$ $\left. R_{\left( 0,2 \right)} \right)$ given by \eqref{AchievableRateMemoryPairsHomog} is achievable for the setting under consideration.
\qed
\end{exmp}

Next, we present the SCC scheme for a general heterogeneous scenario, achieving the memory-rate pair $\left( M_{(p,q)}, R_{(p,q)} \right)$ given by \eqref{AchievableRateMemoryPairsHeter}, for any $p\in \left[ 0:K_w \right]$ and $q \in \left[ p:K_w \right]$. For a given $(p,q)$ pair, where $p \in \left[ 0:K_w \right]$ and $q \in \left[ p:K_w \right]$, each file $W_f$, $f \in [N]$, is divided into $(q-p+1)$ non-overlapping subfiles, represented by
\begin{equation}\label{DivideFilesHeter}
{W_f} = \left( {W_f^{\left( p \right)},...,W_f^{\left( q \right)}} \right),
\end{equation}
where subfile $W_f^{\left( i \right)}$, for $i \in \left[ p:q \right]$, has a rate of
\begin{equation}\label{RateEachSubfileHeter}
{R^{(i)}} \buildrel \Delta \over = \frac{{\gamma \left( {p,{\boldsymbol{\delta}},i} \right)}}{{\sum\limits_{j = p}^q { {\gamma \left( {p,{\boldsymbol{\delta}},j} \right)} } }}R,
\end{equation}
such that $\sum\limits_{i = p}^q {{R^{\left( i \right)}}}  = R$. Note that, $\gamma \left( {p,{\boldsymbol{\delta}},i} \right)$, for $i \in \left[ p:q\right]$, is given in \eqref{gammapiHeter}.

In the sequel, the placement and delivery phases of the SCC scheme are explained, and the achievability of the memory-rate pair in \eqref{AchievableRateMemoryPairsHeter} is proven. 

\subsection{Placement Phase}\label{PlacementPhaseSCC}
In the placement phase, for each set of subfiles $W_1^{(i)}, ..., W_N^{(i)}$ a cache placement procedure, corresponding to the one proposed in \cite[Algorithm 1]{MaddahAliCentralized} for a cache capacity of $iN/K_w$, is performed, $\forall i \in \left[ p:q \right]$; that is, each subfile $W_f^{(i)}$ is partitioned into $\binom{K_w}{i}$ independent equal-rate pieces,
\begin{align}\label{PiecesHeter}
W_f^{\left( i \right)} = &\left( {W_{f,\mathcal{S}^{(i)}_1}^{\left( i \right)},W_{f,\mathcal{S}^{(i)}_2}^{\left( i \right)}, ...,W_{f,\mathcal{S}^{(i)}_{\binom{K_w}{i}}}^{\left( i \right)}} \right),  \forall f \in \left[ N \right], \forall i \in \left[ p:q \right].
\end{align}
The piece $W_{f,\mathcal{S}^{(i)}_l}^{(i)}$ of rate $R^{(i)}/\binom{K_w}{i}$ is cached by receivers $k \in \mathcal{S}^{(i)}_l$, for $l=1, ..., \binom{K_w}{i}$. Thus, the content placed in the cache of each weak receiver $k \in \left[ K_w \right]$ is given by
\begin{equation}\label{CachedContentHeter}
{Z_k} = \bigcup\limits_{f \in \left[ N \right]} {\bigcup\limits_{i \in \left[ {p:q} \right]} {\bigcup\limits_{l \in \left[ \binom{K_w}{i} \right]:k \in {\cal S}_l^{(i)}} {W_{f,{\cal S}_l^{(i)}}^{\left( i \right)}} } }.
\end{equation}
Accordingly, $\binom{K_w-1}{i-1}$ pieces, each of rate $R^{(i)}/\binom{K_w}{i}$, corresponding to each subfile $W_d^{(i)}$ are cached by each weak receiver $k \in \left[ K_w \right]$, which requires a total cache capacity of
\begin{align}\label{CacheCapacityHeter}
M &= N\sum\limits_{i = p}^q {\binom{K_w-1}{i-1}\frac{{{R^{\left( i \right)}}}}{{\binom{K_w}{i}}}} = \frac{N}{K_w}\sum\limits_{i = p}^q {i{R^{\left( i \right)}}} = \frac{N{\sum\limits_{i = p}^q {i \gamma \left( {p,\boldsymbol{\delta},i} \right)} }}{K_w{\sum\limits_{i = p}^q {\gamma \left( {p,\boldsymbol{\delta},i} \right)} }}R.
\end{align}

\subsection{Delivery Phase}\label{DeliveryPhaseSCC}
In the delivery phase, the goal is to satisfy all the demands for an arbitrary demand combination $\left( d_1, ..., d_K \right)$. The delivery phase consists of $(q-p+2)$ different messages, transmitted over orthogonal time periods, where the codewords of the $i$-th message are of length $\beta_i n$ channel uses, for $i=1, ..., q-p+2$, such that $\sum\limits_{i = 1}^{q - p + 2} {{\beta _i}}  = 1$.

The first message of the delivery phase is only targeted for the weak receivers, and the goal is to deliver the missing bits of subfile $W_{d_k}^{(q)}$ to receiver $k$, $\forall k \in [K_w]$. It is to be noted that, for $q=K_w$, based on the cache contents in \eqref{CachedContentHeter}, all the weak receivers have all the subfiles $W_{f}^{(q)}$, $\forall f \in [N]$; therefore, no message needs to be delivered. In the sequel, we consider $q < K_w$. The first message of the delivery phase is transmitted over $\binom{K_w}{q+1}$ orthogonal time slots, where in each slot, a sub-message is delivered to a group of $q+1$ weak receivers. Sub-message $j$ is a codeword of length $\beta _{1,j}n$ channel uses, and is targeted to the receivers in set $\mathcal{S}^{(q+1)}_{j}$, for $j=1, ..., \binom{K_w}{q+1}$, such that $\sum\limits_{j = 1}^{\binom{K_w}{q+1}} {{\beta _{1,j}}}  = \beta _1$. Following the procedure in \cite[Algorithm 1]{MaddahAliCentralized}, the content delivered by sub-message $j$ is given by
\begin{equation}\label{DeliveredContentsHeterPart1}
V_j^{\left( q \right)} \buildrel \Delta \over = {  {\bigoplus}_{k \in {\cal S}_j^{\left( {q + 1} \right)}} W_{{d_k},{\cal S}_j^{(q + 1)}\backslash \{k\}}^{\left( q \right)}}, \quad \mbox{for $j=1, ..., \binom{K_w}{q+1}$}.
\end{equation}
After receiving $V_j^{\left( q \right)}$, each receiver $k \in \mathcal{S}^{(q+1)}_{j}$ can obtain $W_{{d_k},{\cal S}_j^{(q + 1)}\backslash \{k\}}^{\left( q \right)}$, for $j=1, ..., \binom{K_w}{q+1}$, i.e., the missing bits of subfile $W_{d_k}^{(q)}$ of its desired file, having access to $Z_k$ given in \eqref{CachedContentHeter}. Thus, together with its cache content, receiver $k \in [K_w]$ can recover $W_{d_k}^{(q)}$.

The delivery technique performed to transmit messages $2, 3, ..., q-p+1$ follows the same procedure. With the message $q-i+1$ of length $\beta_{q-i+1}n$ channel uses, the server delivers the missing bits of subfile $W_{d_k}^{(i)}$ to each weak receiver $k$, $k \in [K_w]$, and $W_{d_l}^{(i+1)}$ to each strong receiver $l$, $l \in [K_w+1:K]$, for $i=q-1, q-2, ..., p$.\footnote{For example, with the second message, subfile $W_{{d_l}}^{\left( q \right)}$ is delivered to each strong receiver $l \in [K_w+1:K]$, and subfile $W_{{d_k}}^{\left( q-1 \right)}$ to each weak receiver $k \in [K_w]$. With the third message, subfile $W_{{d_l}}^{\left( q-1 \right)}$ is delivered to each strong receiver $l \in [K_w+1:K]$, and subfile $W_{{d_k}}^{\left( q-2 \right)}$ to each weak receiver $k \in [K_w]$, and so on so forth.} Message $(q-i+1)$ is delivered through $\binom{K_w}{i+1}$ sub-messages, transmitted over orthogonal time periods, where sub-message $j$ is of length $\beta_{q-i+1,j}n$ channel uses, such that $\sum\limits_{j = 1}^{\binom{K_w}{i+1}} {{\beta _{q-i+1,j}}}  = \beta _{q-i+1}$. With the $j$-th sub-message, using the coded delivery procedure in \cite[Algorithm 1]{MaddahAliCentralized}, the coded content
\begin{equation}\label{DeliveredContentsHeterPart2}
V_j^{\left( i \right)} \buildrel \Delta \over = {{\bigoplus}_{k \in {\cal S}_j^{\left( {i + 1} \right)}} W_{{d_k},{\cal S}_j^{(i + 1)}\backslash \{k\}}^{\left( i \right)}},
\end{equation}
is delivered to the weak receivers in set ${\cal S}_j^{\left( {i + 1} \right)}$, while 
\begin{equation}\label{DeliveredContentsWeakHeterPart2Period}
\left\{  W_{{d_{K_w+1}},{\cal S}_j^{\left( {i + 1} \right)}}^{\left( i+1 \right)}, ...,  W_{{d_{K}},{\cal S}_j^{\left( {i + 1} \right)}}^{\left( i+1 \right)} \right\},
\end{equation}
is delivered to the strong receivers, for $i=q-1, ..., p$, and $j=1, ..., \binom{K_w}{i+1}$. Observe that, after receiving sub-message $V^{(i)}_j$, each receiver $k \in \mathcal{S}^{(i+1)}_{j}$ can obtain $W_{{d_k},{\cal S}_j^{(i + 1)}\backslash \{k\}}^{\left( i \right)}$, for $j=1, ..., \binom{K_w}{i+1}$, i.e., the missing bits of subfile $W_{d_k}^{(i)}$ of its desired file, for $i=q-1, ..., p$. Note that, the content in \eqref{DeliveredContentsWeakHeterPart2Period}, which is targeted to the strong receivers, is known by each weak receiver in set ${\cal S}_j^{\left( {i + 1} \right)}$. Therefore, the $j$-th sub-message of message $q-i+1$ can be transmitted using joint encoding:
\begin{align}\label{DeliveredCodedContentsPiggybackHeterP1}
{\rm{JE}}\left( {{{\left( V_j^{(i)} \right)}_{\mathcal{S}_j^{(i+1)}}},{{\left( W_{{d_{K_w+1}},{\cal S}_j^{\left( {i + 1} \right)}}^{\left( i+1 \right)}, ..., W_{{d_{K}},{\cal S}_j^{\left( {i + 1} \right)}}^{\left( i+1 \right)} \right)}_{{\left[ K_w+1:K \right]}}}} \right), \quad &\mbox{for $i=q-1, ..., p$,} \nonumber\\
& \quad \; \mbox{and $j=1, ..., \binom{K_w}{i+1}$}.
\end{align}
However, to increase the efficiency of the delivery phase, the $j$-th sub-message is delivered via $K_s$ orthogonal time periods, such that in the $m$-th period a codeword of length $\beta_{q-i+1,j,m}n$ channel uses is transmitted, where $\sum\limits_{m = 1}^{K_s} {{\beta _{q-i+1,j,m}}}  = \beta _{q-i+1,j}$. Coded content $V_j^{(i)}$, targeted for receivers in set $\mathcal{S}_j^{(i+1)}$, is divided into $K_s$ non-overlapping equal-rate pieces
\begin{equation}\label{TildeXDivideHeter}
V_j^{(i)} = \left( {V_{j,1}^{(i)},...,V_{j,{K_s}}^{(i)}} \right),
\end{equation}
and the delivery over the $m$-th time period is performed by joint encoding:
\begin{equation}\label{DeliveredCodedContentsPiggybackHeterP1Sub}
{\rm{JE}}\left( {{{\left( V_{j,m}^{(i)} \right)}_{\mathcal{S}_j^{(i+1)}}},{{\left( W_{{d_{K_w+m}},{\cal S}_j^{\left( {i + 1} \right)}}^{\left( i+1 \right)} \right)}_{{\left\{ K_w+m \right\}}}}} \right), \quad \mbox{for $m=1, ..., K_s$}.
\end{equation}
We note that, after receiving messages 2 to $q-p+1$, each weak receiver $k \in [K_w]$ can obtain subfiles $\left( W^{(q-1)}_{d_k}, W^{(q-2)}_{d_k}, ..., W^{(p)}_{d_k}  \right)$, while each strong receiver $l \in [K_w+1:K]$ can decode subfiles $\left( W^{(q)}_{d_l}, W^{(q-1)}_{d_l}, ..., W^{(p+1)}_{d_l}  \right)$; therefore, together with message 1, the demand of weak receivers are fully satisfied. However, strong receiver $l \in [K_w+1:K]$ only requires to receive subfile $W^{(p)}_{d_l}$.

The last message delivers subfile $W^{(p)}_{d_l}$ to the strong receiver $l \in [K_w+1:K]$ using the channel coding scheme for standard packet erasure BCs.

The contents delivered with each message in the delivery phase for the heterogeneous scenario are summarized in Table \ref{TableDeliveryHeter}.

\begin{table}[!t]
\caption{Contents sent with messages $1$ to $q-p+2$ in the delivery phase of the SCC scheme for the heterogeneous scenario.}
\centering
\begin{tabular}{ ||c||c|c|| }
\hline
\vtop{\hbox{\strut Message}\hbox{\strut \; index}} & \vtop{\hbox{\strut Sub-message}\hbox{\strut \quad \; index}} & Delivered content\\
\hline \hline 
\multirow{4.5}{*}{1} & 1 & $\left( V_1^{\left( q \right)} \right)$ to receivers in ${\cal S}_1^{\left( {q + 1} \right)}$\\
\cline{2-3}
& $ \vdots $ & $\vdots$\\
\cline{2-3}
& $\binom{K_w}{q+1}$ & $\left( V_{\binom{K_w}{q+1}}^{\left( q \right)} \right)$ to receivers in ${\cal S}_{\binom{K_w}{q+1}}^{\left( {q + 1} \right)}$\\
\hline \hline
\multirow{8}{*}{\vtop{\hbox{\strut $t=2, ...,$}\hbox{\strut \;\;\;\;\; $q-p+1$}}} & \multirow{2}{*}{1} & \vtop{\hbox{\strut ${\rm{JE}}\left( {{{\left( V_{1,m}^{(q-t+1)} \right)}_{\mathcal{S}_1^{(q-t+2)}}},{{\left( W_{{d_{K_w+m}},{\cal S}_1^{\left( {q-t+2} \right)}}^{\left( q-t+2 \right)} \right)}_{{\left\{ K_w+m \right\}}}}} \right)$}\hbox{\strut \qquad \qquad \qquad \;\;\;\quad in $m$-th time period, for $m=1, ..., K_s$}}\\
\cline{2-3}
& $ \vdots $ & $\vdots$\\
\cline{2-3}
& \multirow{2.1}{*}{$\binom{K_w}{q-t+2}$} & \vtop{\hbox{\strut ${\rm{JE}}\left( {{{\left( V_{\binom{K_w}{q-t+2},m}^{(q-t+1)} \right)}_{\mathcal{S}_{\binom{K_w}{q-t+2}}^{(q-t+2)}}},{{\left( W_{{d_{K_w+m}},{\cal S}_{\binom{K_w}{q-t+2}}^{\left( {q-t+2} \right)}}^{\left( q-t+2 \right)} \right)}_{{\left\{ K_w+m \right\}}}}} \right)$}\hbox{\strut \qquad \qquad \qquad \;\;\;\quad in $m$-th time period, for $m=1, ..., K_s$}}\\
\hline \hline
$q-p+2$ & \multicolumn{2}{|c||}{$\left( {W_{{d_l}}^{\left( p \right)}} \right)$ to receiver $l \in \left[ K_w+1:K \right]$}\\
\hline
\end{tabular}
\label{TableDeliveryHeter}
\end{table}

\begin{remark}\label{HighDimJoinEncoding}
We remark here that, instead of multicasting XORed contents to groups of weak receivers, one can also use a higher dimensional joint encoding scheme, in which case each weak receiver can decode its missing part directly by using the parts available in its cache as side information. However, we stick to the coding scheme presented here as it makes the connection to the original delivery scheme of Maddah-Ali and Niesen in \cite{MaddahAliCentralized} more explicit.  
\end{remark}

\subsection{Achievable Memory-Rate Pair Analysis (Proof of Theorem \ref{AchievablePoints})}\label{RateAnalysis}
The rate of the coded content targeted to a group of weak receivers for each message of the delivery phase is allocated such that it can be decoded by the weakest receiver among the intended group of weak receivers. 

With sub-message $j$ of message $1$ of length $\beta_{1,j}n$ channel uses, $V_{j}^{(q)}$, which is given in \eqref{DeliveredContentsHeterPart1}, a message of rate $R^{(q)}/\binom{K_w}{q}$ is transmitted to the weak receivers in ${\cal S}_j^{(q + 1)}$, for $j=1, ..., \binom{K_w}{q+1}$. The rate of $V_j^{\left( q \right)}$ is adjusted such that the weakest receiver in $\mathcal{S}^{(q+1)}_{j}$ can decode it successfully, i.e., 
\begin{equation}\label{AchievableDeliveryRateHeterPart11}
\frac{R^{(q)}}{\binom{K_w}{q}} \le {\beta _{1,j}}\left( {1 - \mathop {\max }\limits_{r \in \mathcal{S}^{(q+1)}_{j}} \left\{ {{\delta _r}} \right\}} \right)F, \quad \mbox{for $j=1, ..., \binom{K_w}{q+1}$},
\end{equation}
which after summing over all the sets $\mathcal{S}^{(q+1)}_{j}$, for $j=1, ..., \binom{K_w}{q+1}$, one can obtain
\begin{equation}\label{AchievableDeliveryRateHeterPart12}
\frac{{{{R}^{(q)}}}}{{\binom{K_w}{q} }}\sum\limits_{r = 1}^{{K_w} - q} {\frac{{\binom{ {K_w} - r}{ q} }}{{1 - {\delta _r}}}}  \le \sum\limits_{j = 1}^{\binom{ {K_w}}{ q + 1} } {{{\beta }_{1,j}}}F  = {{\beta }_1}F.
\end{equation}

Note that with the codeword given in \eqref{DeliveredCodedContentsPiggybackHeterP1Sub}, $V_{j,m}^{(i)}$, targeted for the receivers in $\mathcal{S}_j^{(i+1)}$, is of rate $R^{(i)}/\left( K_s \binom{K_w}{i} \right)$, while $W_{{d_{K_w+m}},{\cal S}_j^{\left( {i + 1} \right)}}^{\left( i+1 \right)}$, destined for receiver $K_w+m$, is of rate $R^{(i+1)}/\binom{K_w}{i+1}$, for $m=1, ..., K_s$, $i=q-1, ..., p$ and $j=1, ..., \binom{K_w}{i+1}$. Proposition \ref{CapRegPiggubackErasureBCProp} suggests that the codeword in \eqref{DeliveredCodedContentsPiggybackHeterP1Sub} can be decoded correctly by the intended receivers if
\begin{equation}\label{RateRegionHeterSubSub}
\max\left\{ {\frac{{{{R}^{(i)}}/\left( {{K_s}\binom{K_w}{ i} } \right)}}{{\left( {1 - \mathop {\max }\limits_{r \in \mathcal{S}_j^{\left( {i + 1} \right)}} \left\{ {{\delta _r}} \right\}} \right)F}},\frac{{{{R}^{(i)}}/\left( {{K_s}\binom{K_w}{ i} } \right) + {{R}^{(i + 1)}}/{\binom{K_w}{ i + 1} }}}{{\left( {1 - {\delta _{{K_w} + m}}} \right)F}}} \right\} \le {\beta _{q - i + 1,j,m}}, \mbox{for $m=1, ..., K_s$},
\end{equation}
where the rate of $V_{j,m}^{(i)}$ is limited by the weakest receiver in $\mathcal{S}^{(i+1)}_{j}$, for $i=q-1, ..., p$, and $j=1, ..., \binom{K_w}{i+1}$. By summing up all the $K_s$ inequalities in \eqref{RateRegionHeterSubSub}, we have
\begin{align}\label{RateRegionHeterSubSubSum}
\max \left\{ {\frac{{{{R}^{(i)}}/{\binom{K_w}{ i} }}}{{\left( {1 - \mathop {\max }\limits_{r \in \mathcal{S}_j^{\left( {i + 1} \right)}} \left\{ {{\delta _r}} \right\}} \right)F}},\left( {\frac{{{{R}^{(i)}}}}{{{K_s}\binom{K_w}{ i} }} + \frac{{{{R}^{(i + 1)}}}}{{\binom{K_w}{ i + 1} }}} \right)\sum\limits_{m = 1}^{{K_s}} {\frac{1}{{\left( {1 - {\delta _{{K_w} + m}}} \right)F}}} } \right\} \le {{\beta }_{q - i + 1,j}}, \nonumber\\
\mbox{for $i=q-1, ..., p$, and $j=1, ..., \binom{K_w}{i+1}$}.
\end{align} 
By the choice of \eqref{RateEachSubfileHeter}, and the fact that
\begin{equation}\label{Gammaiplus1Gammai}
\gamma \left( p,\boldsymbol{\delta},i+1 \right) = \frac{\binom{K_w}{i+1}}{\binom{K_w}{i}K_s} \left( \frac{{{K_s}}}{{\left( {1 - {\delta _{{K_w} - i}}} \right)\sum\limits_{l = {K_w} + 1}^K {\frac{1}{{1 - {\delta _l}}}} }} - 1 \right) \gamma \left( p,\boldsymbol{\delta},i \right),
\end{equation}
which follows from the definition in \eqref{gammapiHeter}, the second term of the maximization in \eqref{RateRegionHeterSubSubSum} is reduced to 
\begin{equation}\label{SecondTermMaximization}
\frac{R^{(i)}/\binom{K_w}{i}}{\left( 1-\delta_{K_w-i} \right)F}. 
\end{equation}
Thus, \eqref{RateRegionHeterSubSubSum} is simplified as follows:
\begin{equation}\label{RateRegionHeterSubSubSumSimplified}
\max \left\{ \frac{{{{R}^{(i)}}/{\binom{K_w}{ i} }}}{{\left( {1 - \mathop {\max }\limits_{r \in \mathcal{S}_j^{\left( {i + 1} \right)}} \left\{ {{\delta _r}} \right\}} \right)F}}, \frac{R^{(i)}/\binom{K_w}{i}}{\left( 1-\delta_{K_w-i} \right)F} \right\} \le {{\beta }_{q - i + 1,j}}, \; \; \mbox{for $i=q-1, ..., p$, $j=1, ..., \binom{K_w}{i+1}$}.
\end{equation}
Note that $\left| \mathcal{S}_j^{\left( {i + 1} \right)} \right|=i+1$; hence, for $i=q-1, ..., p$,
\begin{equation}\label{maxdeltaj}
\mathop {\max }\limits_{r \in \mathcal{S}_j^{\left( {i + 1} \right)}} \left\{ {{\delta _r}} \right\} \ge \delta_{K_w-i}, \quad \forall j \in \left[ \binom{K_w}{i+1} \right]. 
\end{equation}
From \eqref{maxdeltaj}, \eqref{RateRegionHeterSubSubSumSimplified} is reduced to
\begin{equation}\label{RateRegionHeterSubSubSumEqui}
\frac{{{{R}^{(i)}}/{\binom{K_w}{ i} }}}{{\left( {1 - \mathop {\max }\limits_{r \in \mathcal{S}_j^{\left( {i + 1} \right)}} \left\{ {{\delta _r}} \right\}} \right)F}}   \le {{\beta }_{q - i + 1,j}}, \quad \mbox{for $i=q-1, ..., p$, $j=1, ..., \binom{K_w}{i+1}$}, 
\end{equation}
which holds for every $j \in \left[ \binom{K_w}{i+1} \right]$, each corresponding to a different $(i+1)$-element subset $\mathcal{S}_j^{\left( {i + 1} \right)}$. After summing up over all values of $j$, one can obtain
\begin{equation}\label{CapacityRegionPiggybackPartiFinalHeter}
\frac{{{{R}^{(i)}}}}{{\binom{K_w}{ i} }}\sum\limits_{r = 1}^{{K_w} - i} {\frac{{\binom{K_w-r}{ i} }}{{1 - {\delta _r}}}}  \le \sum\limits_{j = 1}^{\binom{K_w}{ i + 1} } {{{\beta }_{q - i + 1,j}}}F  = {{\beta }_{q - i + 1}}F, \quad \mbox{for $i=q-1, ..., p$}.
\end{equation}

According to Proposition \ref{CapRegStandardErasureBCProp}, each receiver $l$, $l \in [K_w+1:K]$, can decode subfile $W^{(p)}_{d_k}$ of rate $R^{(p)}$, delivered by the last message, correctly, if
\begin{equation}\label{CapacityRegionPiggybackLastPartHeter}
{{R}^{\left( p \right)}}\sum\limits_{l = {K_w} + 1}^K {\frac{1}{{1 - {\delta _l}}}}  \le {{\beta }_{q - p + 2}}F.
\end{equation}

By combining inequalities \eqref{AchievableDeliveryRateHeterPart12}, \eqref{CapacityRegionPiggybackPartiFinalHeter} and \eqref{CapacityRegionPiggybackLastPartHeter}, we have
\begin{equation}\label{CapacityRegionSCCGeneralHeter}
\sum\limits_{i = p}^q {\left( \frac{{{{R}^{(i)}}}}{{\binom{K_w}{ i} }}\sum\limits_{j = 1}^{{K_w} - i} {\frac{{\binom{K_w-j}{ i} }}{{1 - {\delta _j}}}} \right)} + {{R}^{\left( p \right)}}\sum\limits_{j = {K_w} + 1}^K {\frac{1}{{1 - {\delta _j}}}} \le \sum\limits_{i = p - 1}^q {{{\beta }_{q - i + 1}}} F = F.
\end{equation}
Finally, by replacing $R^{(i)}$, for $i=p,...,q$, with the expression in \eqref{RateEachSubfileHeter}, one can obtain
\begin{equation}\label{CapacityRegionSCCGeneralSimplified2Heter}
R \le \frac{{F\sum\limits_{i = p}^q {\gamma \left( {p,\boldsymbol{\delta},i} \right)} }}{{\sum\limits_{i = p}^q {\left( {\frac{{\gamma \left( {p,\boldsymbol{\delta},i} \right)}}{{\binom{K_w}{i} }}\sum\limits_{j = 1}^{{K_w} - i} {\frac{{\binom{K_w-j}{i} }}{{1 - {\delta _j}}}} } \right)}  + \sum\limits_{j = {K_w} + 1}^{{K}} {\frac{1}{{1 - {\delta _j}}}} }},
\end{equation}
which, together with the cache capacity of each weak receiver, $M$, given in \eqref{CacheCapacityHeter}, proves the achievability of the memory-rate pairs $\left( M_{(p,q)}, R_{(p,q)} \right)$ in \eqref{AchievableRateMemoryPairsHeter}.

\begin{remark}\label{RemarkHomogeneousDeliveredMessages}
We remark that, for the homogeneous scenario, the codeword in \eqref{DeliveredCodedContentsPiggybackHeterP1Sub} can be received by the targeted receivers correctly if
\begin{align}\label{RateRegionHomogSubSubSub}
& \max\left\{ {\frac{{{{R}^{(i)}}/{\left( K_s\binom{K_w}{i} \right)}}}{{\left( {1 - \delta_w} \right)F}},\frac{{{{R}^{(i)}}/{\left( K_s\binom{K_w}{i} \right)} + {{R}^{(i + 1)}}/{\binom{K_w}{ i + 1} }}}{{\left( {1 - {\delta _{s}}} \right)F}}} \right\} \le {\beta _{q - i + 1,j,m}}, \nonumber\\
& \qquad \qquad \qquad \qquad \qquad \qquad \qquad \mbox{for $i=q-1, ..., p$, $j=1, ..., \binom{K_w}{i+1}$, and $m=1, ..., K_s$},
\end{align}
from which we obtain 
\begin{align}\label{RateRegionHomogSubSub}
& \max\left\{ {\frac{{{{R}^{(i)}}}}{\binom{K_w}{ i}{\left( {1 - \delta_w} \right)F}},\frac{{{{R}^{(i)}}/{\binom{K_w}{ i} } + {K_s{R}^{(i + 1)}}/{\binom{K_w}{ i + 1} }}}{{\left( {1 - {\delta _{s}}} \right)F}}} \right\} \le \sum\limits_{m = 1}^{K_s} {{\beta _{q - i + 1,j,m}}} = {\beta _{q - i + 1,j}}, \nonumber\\
& \qquad \qquad \qquad \qquad \qquad \qquad \qquad \qquad \qquad \qquad \mbox{for $i=q-1, ..., p$, and $j=1, ..., \binom{K_w}{i+1}$}.
\end{align}
By adding all $\binom{K_w}{i+1}$ sub-messages, each weak receiver and each strong receiver can decode their targeted contents after receiving message $q-i+1$, if
\begin{equation}\label{CapacityRegionPiggybackParti}
\max \left\{ {\frac{{\frac{{{K_w} - i}}{{i+1}}{R^{(i)}}}}{{\left( {1 - {\delta _w}} \right)F}},\frac{{{\frac{{{K_w} - i}}{{i+1}}{R^{(i)}}} + {K_s}{R^{(i+1)}}}}{{\left( {1 - {\delta _s}} \right)F}}} \right\} \le \sum\limits_{j = 1}^{\binom{K_w}{ i + 1} } {{{\beta }_{q - i + 1,j}}} = {{\beta }_{q - i + 1}}, \quad \mbox{for $i=q-1, ..., p$}. 
\end{equation}
According to \eqref{RateEachSubfileHeter}, we have
\begin{equation}\label{Riplus1RiHomog}
R^{(i+1)}=\frac{\gamma \left( p,\boldsymbol{\delta}_{ws},i+1 \right)}{\gamma \left( p,\boldsymbol{\delta}_{ws},i \right)}R^{(i)} = \frac{K_w-i}{K_s(i+1)} \left( \frac{1-\delta_s}{1-\delta_w} -1 \right) R^{(i)},
\end{equation}
where we used 
\begin{equation}\label{Gammaiplus1GammaiHomog}
\gamma \left( p,\boldsymbol{\delta}_{ws},i+1 \right) = \frac{K_w-i}{K_s(i+1)} \left( \frac{1-\delta_s}{1-\delta_w} -1 \right) \gamma \left( p,\boldsymbol{\delta}_{ws},i \right)
\end{equation}
from \eqref{gammapiHomog}. By substituting \eqref{Riplus1RiHomog} into \eqref{CapacityRegionPiggybackParti}, the second term of the maximization in \eqref{CapacityRegionPiggybackParti} can be rewritten as
\begin{equation}\label{SecondTermEqualityHomog}
\frac{\frac{K_w-i}{i+1}R^{(i)}+\frac{K_w-i}{i+1}\left( \frac{1-\delta_s}{1-\delta_w} -1 \right)R^{(i)}}{\left( 1-\delta_s \right)F} = \frac{\frac{K_w-i}{i+1}R^{(i)}}{\left( 1-\delta_w \right)F},
\end{equation}
which is equal to the first term of the maximization in \eqref{CapacityRegionPiggybackParti}. Thus, \eqref{CapacityRegionPiggybackParti} can be simplified as
\begin{equation}\label{CapacityRegionPiggybackPartiFinal}
\frac{{\frac{{{K_w} - i}}{{i+1}}{R^{(i)}}}}{{\left( {1 - {\delta _w}} \right)F}} \le {\beta _{q-i+1}}, \quad \mbox{for $i=q-1, ..., p$}.
\end{equation}
By combining \eqref{AchievableDeliveryRateHeterPart12}, \eqref{CapacityRegionPiggybackPartiFinal}, and \eqref{CapacityRegionPiggybackLastPartHeter}, we have
\begin{equation}\label{CapacityRegionSCCGeneralSimplified1}
\sum\limits_{i = p}^q {\left( {\frac{{\frac{{{K_w} - i}}{{i + 1}}{R^{(i)}}}}{{\left( {1 - {\delta _w}} \right)F}}} \right) + } \frac{{{K_s}{R^{\left( p \right)}}}}{{\left( {1 - {\delta _s}} \right)F}} \le \sum\limits_{i = p - 1}^q {{{\beta }_{q - i + 1}}} =1.
\end{equation}
Replacing $R^{(i)}$, for $i=p,...,q$, with the expression in \eqref{RateEachSubfileHeter} for the homogeneous scenario results in
\begin{equation}\label{CapacityRegionSCCGeneralSimplified2}
R \le \frac{{F\sum\limits_{i = p}^q {{\gamma \left( {p,\boldsymbol{\delta}_{ws},i} \right)}} }}{{\frac{1}{{1 - {\delta _w}}}\sum\limits_{i = p}^q {\left( {\frac{{{K_w} - i}}{{i + 1}}\gamma \left( {p,\boldsymbol{\delta}_{ws},i} \right)} \right)}  + \frac{{{K_s}}}{{1 - {\delta _s}}}}},
\end{equation}
where $\gamma \left(p,\boldsymbol{\delta}_{ws},i \right)$ is given in \eqref{gammapiHomog}. Observe that \eqref{CapacityRegionSCCGeneralSimplified2} together with the cache capacity of each weak receiver given in \eqref{CacheCapacityHeter} confirm the achievability of the memory-rate pair in \eqref{AchievableRateMemoryPairsHomog} for the homogeneous scenario.
\end{remark}

\begin{remark}
It is to be noted that the achievable STW scheme introduced in \cite{ShirinErasureChannelJournal} is a special case of the SCC scheme for the homogeneous scenario with $q=p+1$. The delivery phase of the STW scheme is performed by delivering three messages in orthogonal time periods. The SCC scheme utilizes a more flexible caching and coding scheme which applies a finer subpacketization compared to \cite{ShirinErasureChannelJournal}, together with the joint encoding scheme of \cite{TuncelPiggyback}, also used in \cite{ShirinErasureChannelJournal}, enabling all the receivers to exploit the cache capacities of the weak receivers.     
\end{remark}

\begin{remark}\label{ErasureProbabilitiesSkewness}
The performance gain from the proposed SCC scheme is relatively higher when the erasure probabilities of the two sets of receivers are more disparate, i.e., the difference between the erasure probability of the best receiver among the weak receivers and the erasure probability of the worst receiver among the strong receivers is larger (see Fig. \ref{Variable_Ks_Kw}). Furthermore, the SCC scheme is relatively robust to skewed erasure probabilities across the strong receivers; this is due to the subpacketization performed to deliver the sub-messages of messages $2$ to $q-p+1$, where the sub-messages targeted to strong receivers are delivered over orthogonal time slots. On the other hand, it suffers more from skewed erasure probabilities across the weak receivers, which is due to the multicast nature of cache-aided coded content delivery to a group of weak receivers, which should be decoded by all of them simultaneously. Accordingly, the reliable rate at which the coded content is delivered to the intended set of weak receivers, is limited by the worst one.         
\end{remark}

\section{Conclusions}\label{Conc}
We have studied cache-enabled content delivery over a packet erasure BC with arbitrary erasure probabilities. We have defined the capacity of this network as the maximum common rate of contents in the library, which allows reliable delivery to all the receivers, independent of their demands. We have derived a lower bound on the capacity by proposing a novel caching and delivery scheme, which enables each receiver, even the strong receivers without a cache memory, to benefit from the cache memories available at the weak receivers. The proposed scheme utilizes a finer subpacketization of the files in the library, and provides a better exploitation of the available cache memories with a higher achievable rate than the state-of-the-art \cite{ShirinErasureChannelJournal}, which focuses on the homogeneous scenario, where the receivers in the same set all have the same erasure probability. This model and the presented results illustrate that even limited storage can be converted into spectral efficiency in noisy communication network, benefiting the whole network, if it is placed strategically across the network, and exploited intelligently.

\bibliographystyle{IEEEtran}
\bibliography{Report}

\end{document}